# A probabilistic mechanics theory for random dynamics

Q.A. Wang

ISMANS, 44 Ave. F.A., Bartholdi, 72000 Le Mans, France

LPEC, UMR 6087, Université du Maine, Ave. O. Messiaen, Le Mans, France

**Abstract**

This is a general description of a probabilistic formalism of mechanics, i.e., an extension of the Newtonian mechanics principles to the systems undergoing random motion. From an analysis of the induction procedure from experimental data to the Newtonian laws, it is shown that the experimental verification of Newton law in a random motion implies a stochastic extension of the virtual work principle and the least action principle, i.e., $\overline{\delta W} = 0$ and $\overline{\delta A} = 0$ averaged over all the random paths instead of $\delta W = 0$ and $\delta A = 0$ for single path in regular dynamics. A probabilistic mechanics is formulated and applied to thermodynamic system. Several known results, rules and principles can be reproduced and justified from this new point of view. To mention some, we have obtain the entropy variation of the free expansion of gas and heat conduction without considering local equilibrium, and a violation of the Liouville theorem and Poincaré recurrence theorem, which allows to relate the entropy production to the work performed by random forces in nonequilibrium process.

## 1) **Background**

This paper gives a general description of a probabilistic mechanics theory for random dynamics. We also outline some previous results [1-7].

The work is more or less inspired by the relationship between classical mechanics and thermodynamics, and the questions concerning the establishment of thermodynamics for nonequilibrium systems. To help the reader to understand the background, the reasoning, the objectives and the methodology, it is useful to give here, without entering at great length into the history and the various points of view concerning nonequilibrium thermodynamics, a brief (surely incomplete) review on the beginning of the story and its actual stage.

The very first persons who tried to relate the second law of thermodynamics to the Newtonian mechanics are Boltzmann, Clausius, Hertz and Helmholtz[1], who have the conviction that the second law is a consequence of the principles of the Newtonian mechanics. But this tentative has as well known failed because the second law is intrinsically in contradiction with the basic principles of mechanics, as shown in the criticisms from Loschmidt, Poincaré[1] and Zermelo[2], and also in the claim of Maxwell with the birth of a small intelligent demon[4]. The key points in the criticisms are the Liouville theorem stipulating the conservation of phase density or volume of Hamiltonian system and the time symmetry of any mechanical dynamics. This story dates back to more than a century ago. The following of the events is well known to the physics community. The key questions concerning the nature and properties of entropy as well as its time evolution never get the last answer. Yet there has been a considerable development of thermodynamics and statistical mechanics for both equilibrium and nonequilibrium systems in the last century. The actual situation can be roughly summarized as following:

- The deduction of macroscopic properties from the microscopic dynamics is rather satisfactory for equilibrium thermodynamics of macroscopic systems. This is a great achievement of the statistical mechanics in the last century, although there are still many open questions around several approaches often used in statistical mechanics for equilibrium states. Some of them are about the physical justification of the variational principles such as maximum entropy principle and maximum entropy production principle (local equilibrium)[9]. For example, is maximum entropy a physical law or just an inference method? Is there some restriction to the functional form of entropy in the principle? Is the Gibbs-Shannon logarithm entropy unique for the variational approach?

- Other questions concerning equilibrium thermodynamics are relative to small size systems such as atomic clusters, biological molecules, nanotube and fullerenes etc, which are definitely different from macroscopic thermodynamic systems in that there are much fewer degrees of freedom and much larger fluctuation effects in their thermodynamic properties. The entire conventional thermodynamics formalism should be carefully used for small systems. Many of the debates around the possible violation of second law by small systems[5][6][7][8] have something to do with the definition of thermodynamic functions and variables such as entropy, internal energy, free energy, work etc.

- There are formulations of thermodynamics[10][11] for nonequilibrium system on the basis of the hypothesis of steady state and local equilibrium in order to use locally the notion of entropy, being defined only for equilibrium system, and to discuss entropy production and related phenomena in nonequilibrium process close to equilibrium.

---

[1] Il résulte de là que les phénomènes irréversibles et le théorème de Clausius ne sont pas explicables au moyen des équation de Lagrange … Toutes les tentatives de cette nature doivent donc être abandonnées[3].

- On the other hand, the bridge between microscopic dynamics and the macroscopic properties of nonequilibrium especially far from equilibrium systems is not yet constructed to date. At least there is no universally accepted answer to the question of increasing property of entropy. The obstacles associated with Liouville theorem, Poincaré recurrence theorem and the time symmetry of microscopic dynamics are always there at least for Hamiltonian system. The main current viewpoints and efforts to solve this question, to mention only some, are the following.

- Boltzmann is correct to define the entropy by $S = k_B \ln \Omega$ and the $H$ function by $H = \sum_i f_i \ln f_i$. The emergence of time arrows from the time symmetrical microscopic dynamics comes essentially from "the great disparity between microscopic and macroscopic scales" and the very large number of degrees of freedom (see for example [13]).

- A widely accepted point of view on the second laws is that it is a probabilistic law, meaning that the likelihood that the law is violated is not zero. This point of view is closely related to the actual debates and reflections around the fluctuation theorems and the possible violations of second law (see for example [5][6][7][8][14][16]). It is nevertheless worthy and instructive to remember that, if one contemplates the crystal clear and simple logic in the discovery of the second law (*no perpetual engine* ➦ *maximal efficiency of reversible engine* ➦ *Kelvin expression of maximal efficiency* ➦ *Clausius entropy and second law*) and the mathematical rigor in this series of deductions from the first assumption about perpetual engine to the discovery of the law, there is indeed no even infinitesimal place for entropy to decrease, in spite the ubiquitous fluctuation in any engine system involved in this logic.

- There has been many suggestions for redefining entropy (functional), entropy production and other thermodynamic quantities for system out of equilibrium with the help of the distributions such as SRB measure (see for example [10][11] [12][15]). It was also suggested by the school of maximum entropy principle to apply this principle - which has been successfully used for equilibrium statistical mechanics and thermodynamic for macroscopic systems - to nonequilibrium system with the Shannon formula as a function of nonequilibrium probability distribution (see for example [2] p69, p107).

- A quite new development of thermodynamics is a combination of conventional thermodynamics and quantum mechanics[17]. In spite of the time symmetry of Schrödinger equation, this tentative is motivated by the facts that the major efforts which failed in deriving macroscopic thermodynamics from microscopic mechanics were classical while the considered constituents (atoms, molecules, clusters …) are all quantum systems.

Many of these viewpoints are in progress. They are not free from the criticism based on the rigorous mathematical reasoning from Liouville theorem, Poincaré recurrence theorem and the time symmetry of microscopic dynamics. Defining second law entropy and entropy production rate for nonequilibrium systems (even they are close to equilibrium) is also a subtle and delicate subject which lacks clear and distinct criterions of validity and risks to encounter exceptions as mentioned in [2] (pp160-167).

To summarize the situation for nonequilibrium statistical mechanics and thermodynamics, it seems still instructive to quote Eddington (see [20], page 77): *"The question whether the second law of thermodynamics and other statistical laws are mathematical deductions from the primary laws, presenting their results in a conveniently usable form, is difficult to answer; but I think it is generally considered that there is an unbridgeable hiatus. At the bottom of all the questions settled by secondary law there is an elusive conception of a priori probability of states of the world which involves an essentially different attitude to knowledge from that presupposed in the construction of the scheme of primary law."*

and

*"It has been a conviction of nearly all physicists that at the root of everything there is a complete scheme of primary law governing the career of every particle or constituent of the world with an iron determinism. This primary scheme is all-sufficing, for, since it fixes the history of every constituent of the world, it fixes the whole world history."*

These remarks are still true until today, despite the introduction of probabilistic and statistical method long ago, despite the irreducible probabilistic character of quantum mechanics: the world of mechanics is intrinsically deterministic. Hamiltonian equations, Lagrange equations, Liouville theorem etc penetrate everywhere in microphysics. This kind of extension from macroscopic to microscopic world is audacious but may be sometimes questionable especially when the laws acquires the character of absolute universal truth, as Poincaré put it[2] [18] : *"It is for apply it when we study the mechanics, and one can apply it only when it remains objective. However, what the principles win in generality and certainty, they lose it in*

---

[2] "Si on étudie la mécanique, c'est pour l'appliquer; et l'on ne peut l'appliquer que si elle reste objective. Or … ce que les principes gagnent en généralité et en certitude, ils le perdent en objectivité. C'est donc surtout avec le côté objectif des principes qu'il convient de se familiariser en bonne heure, et on ne peut le faire qu'en allant du particulier au général, au lieu de suivre la démarche inverse." [18]

*objectivity. Hence it is especially with the objective aspect that one should familiarize himself as early as possible, and one can do that only in going from particular to general, instead of following the inverse approach.* (see [18], p165) "

Our work is a tentative, among many others of courses, to revisit the deterministic mechanics law with this paradigm in mind. The starting point of our approach is best illustrated by a quote from Balian [19]:

"… *every physical law has a probabilistic character: the very principle of scientific knowledge, whatever is the domain, is experience or observation which produces data always with more or less randomness. The laws and predictions result by induction from these data and suffer from the same uncertainty.*"[3]

Our idea is to push this remark farther and to exploit further the experimental uncertainty in reviewing the induction procedure leading to the Newtonian laws of classical mechanics, with the objective to introduce, or to replace the dynamical uncertainty back into the iron determinism which has been, with the induction logic (obviously the most reasonable and scientific way to make rational reasoning from observations), extracted from uncertain experimental data by, inevitably, rejecting the randomness (or accidents) of the motions and keeping only the regularity.

Since it is inevitable to treat dynamical trajectory, the technique of path integral developed in the Feynmann's formulation of quantum mechanics[39] is useful for this framework. The only difference is that the paths here are observed physical object of classical motion while the paths in quantum physics are rather mathematical objects.

2) **Methodology and hypotheses**

The key concern of statistical mechanics is that the macroscopic world is composed of microscopic bodies obeying, according to the wide granted view, reversible dynamics, while the macroscopic dynamics may be irreversible (second law). Since there is no room (it seems to us) for questioning the general validity of the second law (at least for large systems), one should logically and reasonably search for the answer in the mechanics theory. One can think either there is something missing in the extension of the mechanics law to the microscopic mechanical world, or there is something missing in the mechanics law itself in comparison to the real world. Our opinion is the latter. The question is what the missing thing is and how to find it.

---

[3] "Notons d'abord que toute loi physique a un caractère probabiliste : le principe même de la connaissance scientifique, dans quelque domaine que ce soit, est l'expérience ou à défaut l'observation, qui fournissent des données toujours plus ou moins aléatoires. Les lois et les prévisions résultent d'une induction à partir de ces données et souffrent de la même incertitude." [19]

This work is an effort to do that by replacing the Newton's second law in the uncertain experimental environment. The basic idea can be traced as follows. When the experimentalist claims the verification of the Newton's second law by experiment, he simply means that there is a relationship $\overline{\vec{F}}(t) = m\overline{\vec{a}}(t) \pm \vec{\sigma}$ averaged over a large number of measurements with an uncertainty $\sigma$ into which the randomness of the motion is squeezed. The very part in the above equality that interests the experimentalist is $\overline{\vec{F}}(t) = m\overline{\vec{a}}(t)$ instead of $\vec{\sigma}$. However, if he is also interested in the random aspect of the dynamics, there are many way to look into the randomness in much more delicate way. For example, he can trace each motion in order to see into the different random trajectories. We will present a straightforward analysis (see below) showing that the existence of the average law $\overline{\vec{F}}(t) = m\overline{\vec{a}}(t)$ implies a stochastic extension of the virtual work principle and the least action principle which become respectively $\overline{\delta W} = 0$ and $\overline{\delta A} = 0$ averaged over all the paths instead of $\delta W = 0$ and $\delta A = 0$ for single path in the regular dynamics[21][22][23][24][25].

The term 'random motion' is used here in the sense that the motion of a classical system contains a regular and a random part and that, if the randomness vanishes, the system recovers the regular dynamics described by the fundamental principles of mechanics.

The system can be either a simplest mechanical system, i.e., a point mass, or composed of a large number of components. One can consider a Brownian particle or a thermodynamic system such as a gas or a solid composed of a large number of atoms and molecules ("*little particles that move around in perpetual motion, attracting each other when they are a little distance apart, but repelling upon being squeezed into one another.*" [29])

The interactions between the components of the system are supposed to be conservative. The Lennard-Jones potential between the particles in a gas or a condensed matter is an example. This means that if the system is isolated, the energy is conserved just like in an isolated gas or solid which is either in equilibrium or out of equilibrium thermodynamically. At the same time, the system can be under the action of external forces which can be non conservative.

When the system is in contact or interaction with environment, there may be energy exchange through heat transfer and mechanical work. The microscopic forces between the components of the system and the environmental elements in contact are supposed conservative. Hence the global energy (studied system plus environment) can be conserved. This microscopic conservation does not exclude the "dissipation" of macroscopic kinetic energy through friction into heat contained in the global system. But from the microscopic view of the energy of the particles, there is no energy loss.

The above assumptions cover a large number of situations in nature. In case where there is no external dissipation forces (friction for example), the system can be *microscopically* considered as Hamiltonian system whose Hamiltonian is given by the sum of all the kinetic and potential energies of the constituents at any moment of time. The dissipation of macroscopic mechanical energy into heat can nevertheless take place when there are macroscopic transport phenomena or macroscopic frictions inside the system and on the surfaces of the system.

This approach is somewhat phenomenological and based on the observation of the multitude of trajectories of random motion and on the probabilistic description without entering into the study of the cause (noises) of the randomness. The dynamics is supposed fundamentally random and unpredictable, meaning that the source of the randomness is objective such as thermal, quantum, chaotic and initially conditional, as discussed in by M. Gell-Mann in [31][32]. Other sources of unpredictability related to the human ignorance, incapability of perception and of mathematical or numerical treatments are not considered. The randomness can come from either internal and intrinsic dynamics of the system or external perturbation such as the molecule noise around a Brownian particle or the random uncontrollable perturbations suffered by any macroscopic system in motion (a falling light body in the air or a driven car on the road, for example).

In what follows, we will begin by outlining the fundamental principles of classical mechanics which are then extended to random motion by considering the non uniqueness of trajectory between two given points in configuration space. The consequences concerning thermodynamics and statistical mechanics are discussed in detail for both equilibrium and nonequilibrium states.

3) Principle of least action

The least action principle (LAP)[4] [30][33] is well formulated for Hamiltonian system (without dissipation) satisfying following equations:

$$\dot{x}_k = \frac{\partial H}{\partial P_k} \text{ and } \dot{P}_k = -\frac{\partial H}{\partial x_k} \text{ with } k=1,2,\ldots g \quad (1)$$

where $x_k$ is the coordinates, $P_k$ the momentum, $H(\dot{x}, P, t)$ the Hamiltonian given by $H = K + V$, $K$ the kinetic energy, $V$ the potential energy, and $g$ the number of degrees of freedom of the system. The Lagrangian is defined by $L = K - V$.

LAP stipulates that the action of a motion between two point $a$ and $b$ in the configuration space defined by the time integral $A = \int_a^b L dt$ on a given path from $a$ to $b$ must be a stationary on the unique true path for a given period of time $\tau$ of the motion, i.e.,

$$\delta A|_\tau = 0 . \quad (2)$$

In what follows, we will drop the index $\tau$ of the variation and the action variation is always calculated for fixed period of time $\tau$. This principle yield the Euler-Lagrange equation given by

$$\frac{\partial}{\partial t}\frac{\partial L}{\partial \dot{x}_k} - \frac{\partial L}{\partial x_k} = 0 \quad (3)$$

with $\dot{x}_k = \frac{\partial x_k}{\partial t}$. The above equation underlies a completely deterministic dynamics: there is only one path between two given points so that all the states of the system are completely determined by Eq.(3) for every moment of the motion.

4) Principle of virtual work

In mechanics, a virtual displacement of a system is a kind of hypothetical infinitesimal displacement with no time passage and no influence on the forces. It should be perpendicular to the constraint forces. The principle of virtual work[34][35] says that the total work performed by all the forces acting on a system in static equilibrium is zero for any possible virtual displacement. Let us suppose a simple case of a system of $N$ points of mass in equilibrium under the action of the forces $F_i$ ($i=1,2,…N$) on point $i$, and imagine virtual displacement of each point $\delta \tilde{r}_i$. According to the principle, the virtual work $\delta W$ of all the forces $F_i$ on all $\delta \tilde{r}_i$ vanishes for static equilibrium, i.e.

---

[4] We continue to use the term "least action principle" here considering its popularity in the scientific community. We know nowadays that the term "optimal action" is more suitable because the action of a mechanical system can have a maximum, or a minimum, or a stationary for real paths[33].

$$\delta W = \sum_{i=1}^{N} \vec{F}_i \cdot \delta \tilde{r}_i = 0 \qquad (4)$$

This principle for static equilibrium problem was extended to "dynamical equilibrium" by d'Alembert[35] who added the inertial force $-m_i \vec{a}_i$ on each point of the system in motion

$$\delta W = \sum_{i=1}^{N} (\vec{F}_i - m_i \vec{a}_i) \cdot \delta \tilde{r}_i = 0 \qquad (5)$$

where $m_i$ is the mass of the point $i$ and $\vec{a}_i$ its acceleration. From this principle, we can not only derive Newtonian equation of dynamics, but also other fundamental principles such as least action principle.

The deterministic character and the uniqueness of trajectory of the dynamics dictated by these two principles can be illustrated in both configuration and phase spaces as shown in figure 1 which tells us that a motion from a point $a$ in configuration space must arrive at point $b$ when the duration of motion ($t_b$-$t_a$) is given. Equivalently in phase space, once the initial point (condition) $a$ is given, the path is then determined, meaning that the unique destination after ($t_b$-$t_a$) is $b$.

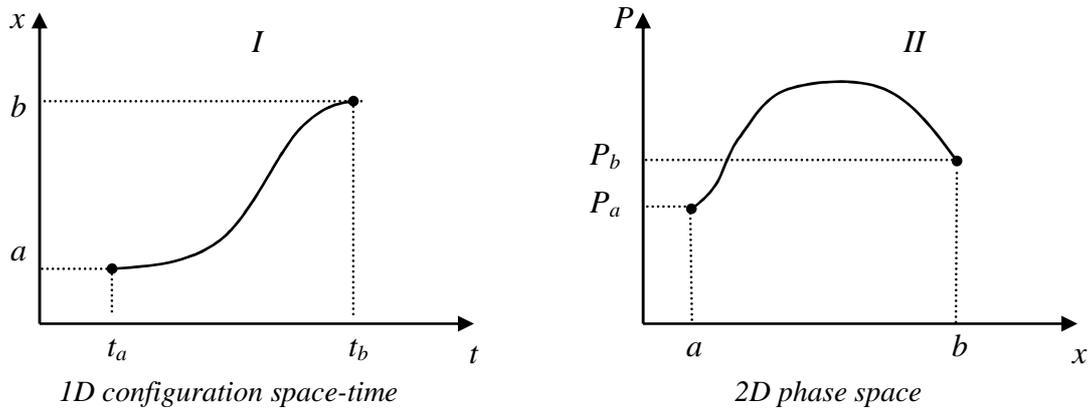

Figure 1: Illustration of a least action path of regular motion of Hamiltonian system between two points $a$ and $b$ in configuration space (I) and in phase space (II). The virtual work on each point of this path is zero according to Eq.(5). The duration of motion ($t_b$-$t_a$) for the path in configuration space is given, while for the phase space path

the duration of motion is not specified since it is hinted in the initial or final conditions (positions and velocities). The meaning of this is that a motion from a given phase point *a* must have a single destination *b*.

## 5) Liouville theorem

Since the Liouville theorem is often involved in the discussions relative to thermodynamic entropy in statistical mechanics, we give an outline of it here and will come back to this topic within the random dynamics.

We look at the time change of phase point density $\rho(x,P,t)$ in a, say, 2-dimensional phase space $\Gamma$ when the system of interest moves on the geodesic, i.e., the path of least action. The time evolution neither creates nor destroys state points, so the number of the phase points is conserved. We then have the conservation law in the phase space

$$\frac{\partial \rho}{\partial t} + \frac{\partial (\dot{x}\rho)}{\partial x} + \frac{\partial (\dot{P}\rho)}{\partial P} = 0 \tag{6}$$

which means

$$\frac{d\rho}{dt} = \frac{\partial \rho}{\partial t} + \frac{\partial \rho}{\partial x}\dot{x} + \frac{\partial \rho}{\partial P}\dot{P} = -\left(\frac{\partial \dot{x}}{\partial x} + \frac{\partial \dot{P}}{\partial P}\right)\rho. \tag{7}$$

For the least action path satisfying Eq.(1), the right hand side of the above equation is zero, leading to the Liouville theorem:

$$\frac{d\rho}{dt} = 0, \tag{8}$$

i.e., the state density in phase space is a constant of motion. The phase volume $\Omega$ available to the system can be calculated by $\Omega = \int_{\Gamma} \frac{1}{\rho} dn$ where $dn$ is the number of phase point in an elementary volume $d\Gamma$ at some point in phase space. The time evolution of the phase volume $\Omega$ accessible to the system is then given by

$$\frac{d\Omega}{dt} = \frac{d}{dt}\int_{\Gamma}\frac{1}{\rho}dn = -\int_{\Gamma}\frac{1}{\rho^2}\frac{d\rho}{dt}dn = 0 \tag{9}$$

meaning that this phase volume is a constant of motion.

The second law of thermodynamics states that the entropy of an isolated system increases or remains constant in time. But the Liouville theorem implies that if the motion of the system obeys the fundamental laws of mechanics, the Boltzmann entropy defined by $S = \ln\Omega$ must be constant in time. On the other hand, if the probability distribution of states $p(x,P)$ in phase space is proportional to $\rho(x,P)$ and that an entropy is a functional of $p(x,P)$, i.e., $S(p)$, then this entropy must be constant in time, which is in contradiction with the second law.

6) Poincaré recurrence theorem

One of the important consequence of the Liouville theorem is the Poincaré recurrence theorem for mechanical system whose total energy is finite[42][43]. This theorem states that almost every point is recurrent in phase space.

It will be useful to appreciate it in more mathematical language. Suppose an initial finite phase volume $\Omega$ accessible to a system, let $x \in \Omega$ be the set of points that is not recurrent and $f^n(x)$ is the positions of this non recurrent set of points after a period of time $t=n$ with a volume $\Omega_n$. Since these points are not recurrent, we must have $f^n(x) \notin \Omega_0$ for any $n=1,2,3$ … with $\Omega_0 = \Omega_1 = \Omega_2 = ... = \Omega_n$ by virtue of the Liouville theorem. Since the trajectories of regular mechanical motion do not intersect between them[33] and $x$ is not recurrent (never returns to $\Omega_0$), the intersection between the sets $\Omega_0, \Omega_1, \Omega_2 ... \Omega_n$ must be empty, meaning that the cumulate volume $\Omega_{cum} = n\Omega_0$ is increasing with time. Then we can always find a value $N$ such that $\Omega_{cum} = n\Omega_0 > \Omega$ for all $n \geq N$. This violates the Liouville theorem of phase conservation, which means that the set of those points $x$ having volume $\Omega_0$ has zero measure ($\Omega_0 = 0$) and that, as a consequence, almost every point of $\Omega$ returns to $\Omega$.

Since the entropy is determined by the ensemble of these points (positions and velocities), it must also return to its initial value. It was Zermelo[44] who pointed out this apparent incompatibility between Poincaré recurrence theorem, the second law, and Boltzmann's H-theorem, in the sense that either the second law may be violated or connection to mechanical system cannot be made. Hence the science of irreversible processes, thermodynamics, could not be reduced to mechanics, and Hertz's mechanical derivation of the second law must be impossible.

We would like to recall that the key element in the proof of this theorem is the conservation of phase volume $\Omega$ of mechanical system stated by Liouville theorem[42][43].

7) Random dynamics

The above mentioned principles hold whenever the motion is regular. In other words, we can refer to any motion which can be described analytically and explicitly by Newtonian laws as regular motion.

On the contrary, we define an irregular or random motion as a dynamics which violates the above paradigms. One of the most remarkable characteristics of random motions is the non uniqueness of paths between two given points in configuration space as well as in phase space. This behavior implies the occurrence of multiple paths to different destinations from a given phase point, which is illustrated in Figure 2.

The cause of the randomness is without any doubt the noises or random forces in and around the observed system. In this work we do not run into the study of the origin of the noises. We only look into the effects, i.e., the multiplicity of paths mentioned above for a motion during a time period, or the multiplicity of states for a given moment of time of the motion.

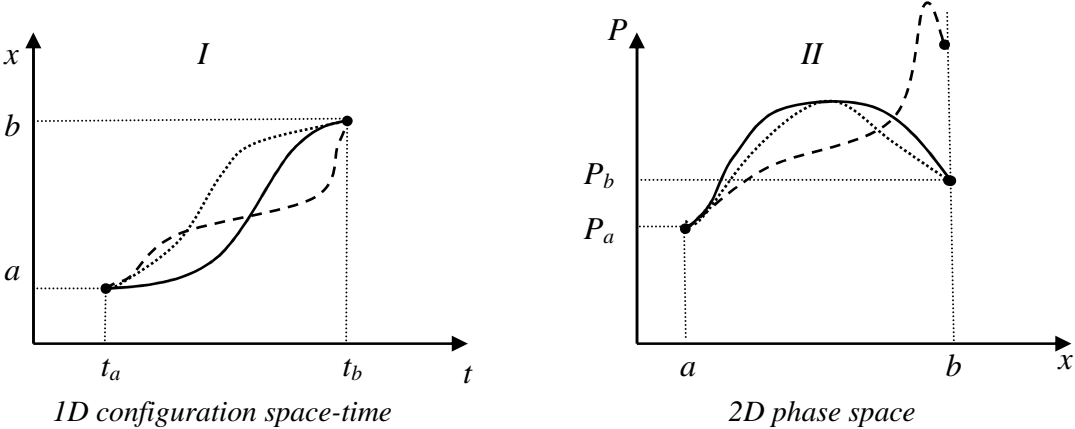

Figure 2: An illustration of the non uniqueness of trajectory of random dynamics. Notice that the three path examples between *a* and *b* in configuration space may have different end points in phase space even if they have the same initial state.

In what follows, we will first introduce the extension of the above fundamental principles to random dynamics by considering a very common case: the descent of a body from an inclined smooth but irregular long surface.

## 8) Observation of a random motion

We imagine an experience to verify Newtonian laws by the descent of a body on an inclined surface. The friction can be neglected although the surface is somewhat rugged. The experience is repeated many times. We look at all the motions of the body starting from a point *a* and arriving at a point *b*.

The outcome of the experiment is the verification of the Newtonian law $\vec{F} = m\vec{a}$ where $\vec{F}$ is the total force on the body, $m$ is its mass and $\vec{a}$ its acceleration. This means that, after a large number of experiences, the means of the experimental data of position, velocity and acceleration agree with the prediction from the Newtonian law. For the sake of simplicity, we suppose that there is no uncertainty in the measure of mass and time, and that the precision of the measure is high enough to see the irregularity of the motion and its deviation from the ideal descending motion on a perfectly smooth and regular surface. This ensures that the uncertainty in the measured results only comes from the irregularity of the surface. The motions are illustrated in Figure 3.

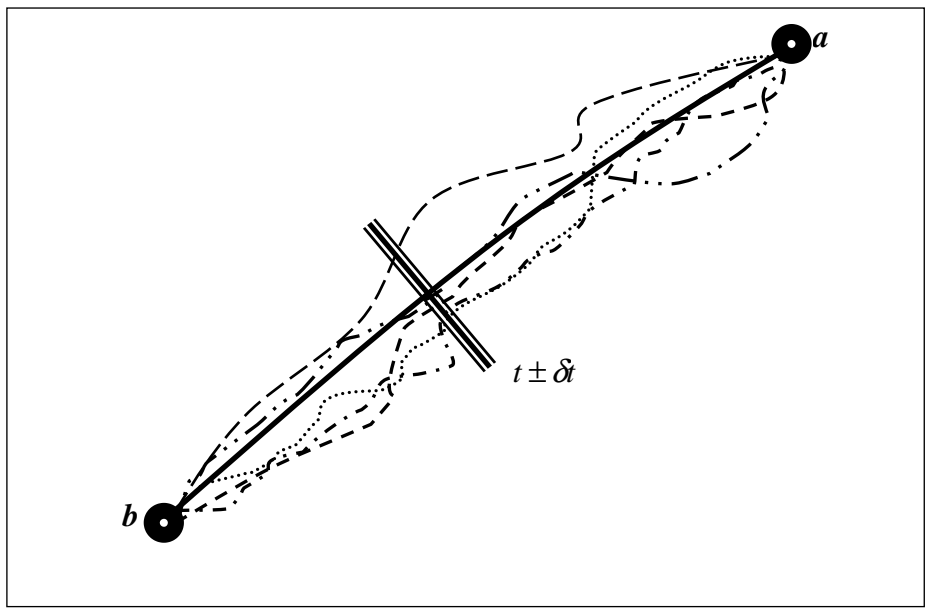

Figure 3: Illustration of the descending motion of a body on an irregular but smooth surface from a point *a* to a point *b*. According to our assumptions, the multiplicity of paths is mainly the consequence of the irregularity of the ramp (not depicted here). The position, velocity and acceleration are measured at the moments of $t$, $t + \delta t$ and $t - \delta t$ marked by the three bars, respectively. The intersections of the paths with the thick bar marking the moment *t* are the possible positions of the body at that moment. The full line linking *a* and *b* is the mean path formed by the successive average positions $\bar{\vec{r}}(t)$ between *a* and *b*. It is along this path that the Newtonian law is verified in the experiment as analyzed in the text. Hence this path is expected to be the unique least action path when the surface is perfectly regular and smooth.

Now let us see in detail a usual experimental method to verify $\vec{F} = m\vec{a}$ by measuring the force and the acceleration at a given moment of time *t*. A large number of measurements will be carried out for a given moment $t$ of the motion in the small time interval from $t$ to $t+\delta t$. The average positions are given by $\bar{\vec{r}}(t) = \frac{1}{M}\sum_{i=1}^{N}\vec{r}_i(t)$ and $\bar{\vec{r}}(t+\delta t) = \frac{1}{M}\sum_{i=1}^{M}\vec{r}_i(t+\delta t)$ where $M$ is the total number of measurements and $i$ is summed over all the measured values of position $\vec{r}_i(t)$. The average velocity is then given by $\bar{\vec{v}}(t) = \frac{\delta \bar{\vec{r}}(t)}{\delta t} = \frac{1}{M}\sum_{i=1}^{M}\frac{\vec{r}_i(t+\delta t)-\vec{r}_i(t)}{\delta t} = \frac{1}{M}\sum_{i=1}^{M}\vec{v}_i(t)$. The same can be done at another moment of time very close to $t$, e.g., $t-\delta t$, to obtain $\bar{\vec{v}}(t-\delta t) = \frac{1}{M}\sum_{i=1}^{M}\vec{v}_i(t-\delta t)$. Then the average acceleration can be calculated by $\bar{\vec{a}}(t) = \frac{1}{M}\sum_{i=1}^{M}\frac{\vec{v}_i(t)-\vec{v}_i(t-\delta t)}{\delta t} = \frac{1}{M}\sum_{i=1}^{M}\vec{a}_i(t)$. The force on the body may depend on its position, i.e., $\vec{F}_i = \vec{F}(\vec{r}_i)$ for $i^{th}$ outcome of the experiment. The mean force at the moment $t$ is given by $\bar{\vec{F}}(t) = \frac{1}{M}\sum_{i=1}^{M}\vec{F}_i(t)$.

If finally the verification is successful, we get

$$\bar{\vec{F}}(t)/m = \bar{\vec{a}}(t) \pm \vec{\sigma} \qquad (10)$$

with some uncertainty characterized by a standard deviation $\vec{\sigma}$.

## 9) Virtual work principle for random dynamics

The full line linking $a$ and $b$ in Figure 3 represents the mean path formed by the successive average positions $\bar{\vec{r}}(t)$ between $a$ and $b$ ($t_a \leq t \leq t_b$). It is along this ideal path (when there is no randomness) that the Newtonian law is verified. Hence this line should be the unique least action path if the surface was perfectly regular and smooth. The virtual work principle Eq.(5) holds on any point of this path and is given here by (let $N=1$)

$$\overline{\delta W} = (\bar{\vec{F}} - m\bar{\vec{a}})\cdot \delta\vec{r} = \frac{1}{M}\sum_{i=1}^{M}(\vec{F}_i - \vec{a}_i)\cdot \delta\vec{r} = 0 \qquad (11)$$

The bar over the virtual work implies that the work is a mean. Let $\delta W_i$ be the virtual work at the measured position $i$ of the $i^{th}$ experience (one of the intersections of the paths with the thick bar at time $t$ in Figure 3, for example), i.e., $\delta W_i = (\vec{F}_i - m\vec{a}_i)\cdot \delta\vec{r}$, Eq.(11) reads:

$$\overline{\delta W} = \frac{1}{M}\sum_{i=1}^{M}\delta W_i = 0. \qquad (12)$$

The most important meaning of Eq.(12) is that *when the Newtonian law is verified as a mean in the random dynamics, the virtual work principle may be violated by individual random motion or events*, i.e., $\delta W_i \neq 0$ *at the position* $\vec{r}_i$*, while it must hold for the average virtual work calculated over the ensemble of events (positions) at a given moment.*

The sum in Eq.(12) over the different experimental outputs at time *t* can be replaced by a sum over the positions $\vec{r}_j(t)$ observed at the same time with $1 \leq j \leq w$ where *w* is the total number of observed positions at *t*. Let $M_j$ be the number of measures getting the output $\vec{r}_j(t)$, Eq. (12) becomes

$$\overline{\delta W} = \sum_{j=1}^{w} \frac{M_j}{M} \delta W_j = \sum_{j=1}^{w} p_j \delta W_j = 0. \tag{13}$$

where $p_j$ is the probability to find the body at $\vec{r}_j(t)$ when *M* is very large. In this expression, we can also choose *j* as the index of the observed states (positions and velocities) of the body. In this case, $p_j$ will be the probability distribution of states.

Eq.(13) will be used as the new virtual work principle for random dynamics. It can be stipulated as follows.

*The statistical mean of the total virtual work done by all the forces acting on a system (equilibrium or not) in random motion must be zero, where the means is taken over all the possible positions or states of the system at a given moment of time.*

Note that, according to the above discussion, the validity of this principle is subject to the condition that the random dynamics verifies the Newtonian laws in a statistical way. However, when this principle will be applied to random systems for which we can not say a priori whether or not the Newtonian laws are verified statistically, the legitimacy of the application should be subject to experimental verification a posteriori.

10) Equilibrium thermodynamics and maximum entropy approach

*a) Virtual work and entropy*

In this section, it will be shown that the well known maximum entropy approach to statistical thermodynamics can be derived from the extended virtual work principle.

Let us consider an ensemble of *equilibrium* systems, each composed of *N* particles in random motion with $\vec{v}_i$ the velocity of the particle *i*. Without loss of generality, the systems are supposed to have no macroscopic motion, i.e., $\sum_{i=1}^{N}\vec{v}_i=0$ at any time. We imagine that a system in thermodynamic equilibrium leaves its equilibrium state by a reversible infinitesimal virtual process. Let $\vec{F}_i$ be the force on the particle $i$ at that moment. $\vec{F}_i$ includes all the interacting forces between the particle $i$ and other particles as well as the walls of the container. In the virtual process, each particle with acceleration $\ddot{\vec{r}}_i$ has a virtual displacement $\delta \vec{r}_i$. The total virtual work over all the displacements is given by

$$\delta W = \sum_{i=1}^{N}(\vec{F}_i - m_i\ddot{\vec{r}}_i) \cdot \delta \vec{r}_i \tag{14}$$

For a given particle $i$ of mass $m_i$, we have $m_i\ddot{\vec{r}}_i \cdot \delta \vec{r}_i = m_i\delta \dot{\vec{r}}_i \cdot \dot{\vec{r}}_i = \delta(\frac{1}{2}m_i\dot{\vec{r}}_i^2) = \delta e_{ki}$ which is the virtual variation of the kinetic energy $e_{ki}$ of the particle.

As we have supposed, there is no dissipative force in the system or on the particles. This implies the existence of the particle-particle or particle-wall potential energy. Let $\vec{F}_i = -\nabla e_{pi}$ with potential energy $e_{pi}$ of the particle $i$. We have

$$\sum_{i=1}^{N}\vec{F}_i \cdot \delta \vec{r}_i = \sum_{i=1}^{N}(-\nabla_i e_{pi}) \cdot \delta \vec{r}_i = -\sum_{i=1}^{N}\delta e_{pi} \tag{15}$$

where $\delta e_{pi} = \nabla e_{pi} \cdot \delta \vec{r}_i$ is the virtual change in potential energy. It follows that

$$\delta W = -\sum_{i=1}^{N}(\delta e_{pi} + \delta e_{ki}) = -\sum_{i=1}^{N}\delta e_i = -\delta E \tag{16}$$

where $\delta e_i = \delta e_{pi} + \delta e_{ki}$ is the virtual variation of the total energy of the particle $i$, and $E = \sum_{i=1}^{N}e_i$ is the total energy of the system.

At this stage, no statistics has been done. The particles are treated as if they were in regular motion. As a matter of fact, the dynamics is random. The randomness is taken into account by the consideration of different possible microstates $j$ with different probability $p_j$ ($j=1,2 \ldots w$) at that moment of virtual variation. If we look at the system in the phase space, the considered virtual process with given virtual displacements may take place on different microstates. Hence Eq.(16) is only the virtual work of the system at one microstate $j$. In view of the fact that the virtual displacement of each particle is given and does not have influence on the forces, this virtual work depends on the microstates through forces and potential energy. It should be written as $\delta W_j = -\sum_{i=1}^{N}(\delta \varepsilon_i)_j$. The mean of this individual virtual work over all the microstates is

$$\overline{\delta W} = \sum_{j=1}^{w} p_j \delta W_j. \tag{17}$$

The virtual work at the state $j$ can be analyzed in more detail as follows. A microstate $j$ is some distribution of the $N$ particles over the one particle states $k$ with energy $\varepsilon_k$ where $k$ varies from, say, 1 to $g$ ($g$ can be very large and tend to infinity for continuous energy spectra). We imagine $N_j$ identical particles distributed over the $g$ states at a microstate $j$ which is here a combination of $g$ numbers $n_k$ of particles over the $g$ states, i.e., $j=\{n_1, n_2, \ldots n_g\}$. We naturally have $N_j = \sum_{k=1}^{g}(n_k)_j$ and $E_j = \sum_{k=1}^{g}(n_k)_j \varepsilon_k$. Since the virtual work only affects the energy of the particle, it can be written as

$$\delta W_j = -\sum_k (n_k)_j \delta \varepsilon_k = -\delta \sum_k (n_k)_j \varepsilon_k + \sum_k (\delta n_k)_j \varepsilon_k \tag{18}$$
$$= -\delta E_j + \sum_k (\delta n_k)_j \varepsilon_k.$$

The first term of the right hand side is the total energy variation due to the one particle energy variation $\delta \varepsilon_k$ caused by the virtual work as well as to the variation in particle number $\delta N_j$ of the system. The second term is just the energy variation caused by the particle number variation $\delta N_j = \sum_k (\delta n_k)_j$. Hence Eq.(17) reads

$$\overline{\delta W} = -\sum_j p_j \delta E_j + \sum_j p_j \sum_k \varepsilon_k (\delta n_k)_j = -\overline{\delta E} + \sum_k \varepsilon_k \overline{\delta n_k} \tag{19}$$
$$= -\overline{\delta E} + \mu \overline{\delta N}$$

where the chemical potential is given by $\mu = \sum_k \varepsilon_k \overline{\delta n_k} / \overline{\delta N}$ with $\overline{\delta N} = \sum_j p_j \delta N_j = \sum_j p_j \sum_k (\delta n_k)_j = \sum_k \overline{\delta n_k}$ and $\overline{\delta E} = \sum_j p_j \delta E_j$. Since $\overline{\delta E} = \delta \overline{E} - \sum_j E_j \delta p_j$ and $\overline{\delta N} = \delta \overline{N} - \sum_j N_j \delta p_j$ with $\overline{E} = \sum_j p_j E_j$ and $\overline{N} = \sum_j p_j N_j$, we get

$$\overline{\delta W} = -\delta \overline{E} + \sum_j E_j \delta p_j + \mu \delta \overline{N} - \mu \sum_j N_j \delta p_j \qquad (20)$$

$$= -\delta \overline{E} + \mu \delta \overline{N} + \sum_j (E_j - \mu N_j) \delta p_j$$

To ensure the conservation of energy of isolated system in equilibrium, the mean virtual work of random forces must vanish for equilibrium state. Comparing then Eq.(20) to the first law $\delta \overline{E} = \delta Q - \delta W + \mu \delta \overline{N}$ for Grand-canonical ensemble, we identify the heat transfer $\delta Q = \sum_j (E_j - \mu N_j) \delta p_j$. For a reversible virtual process, we can write

$$\delta S = \beta \delta Q = \beta \sum_j (E_j - \mu N_j) \delta p_j \qquad (21)$$

and get

$$\overline{\delta W} = -\delta \overline{E} + \mu \delta \overline{N} + \frac{\delta S}{\beta}. \qquad (22)$$

where $S$ is the thermodynamic entropy of the second law in view of the Eq.(21).

## b) *Extremum entropy algorithm*

The extended virtual work principle yields

$$\delta(S - \beta \overline{E} + \beta \mu \overline{N}) = 0 \qquad (23)$$

which can be called extremum entropy approach for grand-canonical ensemble with the energy and particle number constraints. We will prove a bit later that this is actually the usual maximum entropy principle.

Notice that here the "constraints" associated with energy and particle number naturally appear in the variational calculus as a simple consequence of virtual work principle, in contrast to the rather subjective character of the introduction of these constraints within the inference theory or inferential statistical mechanics[36] by the arguments such as that the averaged value of an observable quantity represents a factual information to be put into the maximization of information in order to derive least biased probability distribution[37].

We can suppose the entropy is a function of the probability distribution $p_j$, i.e., $S = f(p_1, p_2, ... p_j ...)$, and write $\delta S = \sum_j \frac{\partial f}{\partial p_j} \delta p_j$ where $\delta p_j$ is the variation of the probability due to the virtual process. Compare this with $\delta S = \beta \sum_j (E_j - \mu N_j) \delta p_j$, one gets

$$\sum_j (\frac{\partial f}{\partial p_j} - \beta E_j + \beta \mu N_j) \delta p_j = 0. \tag{24}$$

Using normalization condition $\sum_j \delta p_j = 0$, one can prove

$$\frac{\partial f}{\partial p_j} - \beta E_j + \beta \mu N_j = \alpha. \tag{25}$$

with a constant $\alpha$. Eq.(25) can be used for deriving the probability distribution of the nonequilibrium component of the dynamics if the functional $f$ is given. Inversely, if the probability distribution is known, one can derive the functional of $S$. It is common knowledge that the Shannon formula for thermodynamic entropy can be derived from the exponential distribution of energy of the Boltzmann statistical mechanics (see also the discussion in the following subsection $c$).

For canonical ensemble, we have $\delta \overline{N} = 0$ and $\delta(S - \beta \overline{E}) = 0$ and $\frac{\partial f}{\partial p_j} - \beta E_j = \alpha$. For microcanonical ensemble, the system is completely closed and isolated with constant energy $\delta \overline{E} = 0$ and constant particle number $\delta \overline{N} = 0$. When the virtual displacements occur, the total virtual work would be transformed into virtual heat such that Eq.(22) becomes $\overline{\delta W} - \overline{\delta Q} = 0$. Then $\overline{\delta W} = 0$ leads to

$$\delta S = 0 \text{ or } \frac{\partial f}{\partial p_j} = \alpha \tag{26}$$

which necessarily yields uniform probability distribution over the different microstates $j$, i.e., $p_j = 1/w$. Note that here the uniform distribution over the microstates is not an a priori assumption, but a direct consequence of virtual work principle.

This equiprobability can be proven as follows only by supposing that $S = f(p_1, p_2, ... p_j ...)$ is a strictly increasing or decreasing function of all $p_j$ throughout the interval $0 \leq p_j \leq 1$, i.e., its derivatives $\frac{\partial f}{\partial p_j} > 0$ or $< 0$ and are zero only at some finite number of points in that interval. However, Eq.(26) tells us that $\frac{\partial f}{\partial p_j} = \alpha$ is a constant independent of $p_j$, implying that $S = f(p_1, p_2, ... p_j ...)$ is either a linear function of all $p_j$, or all $p_j$ are identical. It is evident that entropy cannot be linear function of $p_j$. The equal probability of all microstates follows.

The main conclusion of this section is that, at thermodynamic equilibrium, maximum entropy under certain constraints is a consequence of the virtual work principle. From Eq.(19), one notices that maximum entropy can be written in the following concise form for any ensemble having $n$ random variables $X_i$ ($i=1,2 \ldots n$):

$$\sum_{i=1}^{n} \chi_i \overline{\delta X_i} = 0 \tag{27}$$

where $\chi_i$ is some constant corresponding to $X_i$. For grand-canonical ensemble, this is $\overline{\delta E} - \mu \overline{\delta N} = 0$ and for canonical ensemble, it is $\overline{\delta E} = 0$.

We would like to mention here that this derivation of maximum entropy is not associated with any given form of entropy, unlike the original version of Jaynes principle.

### c) Maximum entropy algorithm

At this stage, it is worth mentioning that the virtual work principle only stipulates vanishing virtual work and that it only yields extremum (maximum/minimum/stationary) of any quantity proportional to virtual work such as the entropy $S$ in this work. In order to see the maximum of $S$ with respect to probability variation, one should see its variational expression $\delta S = \beta \sum_j E_j \delta p_j$ (we take canonical ensemble for simplicity). Since $p_j$ is the probability distribution of the random variable $E_j$, i.e., $p_j = f(E_j)$, we have $\delta S = \beta \sum_j f^{-1}(p_j) \delta p_j$ or by integration for strict monotonic $f(E_j)$ :

$$S = \beta \sum_j \int_0^{p_j} f^{-1}(p) dp - S_0 \tag{28}$$

where $S_0 = \int_0^{p_j} f^{-1}(p)dp$ in order that $S$ vanish whenever a $p_j$ is equal to one for non probabilistic case or regular dynamics. This means that if $f(E_j)$ is a strict decreasing function with positive $\beta$, $S$ will be concave and Eq.(23) is a maximum entropy. This is the case of the Boltzmann exponential distribution with positive temperature. However, if $f(E_j)$ is strictly increasing, $S$ will be convex and negative according to Eq.(28) with positive $\beta$. In this case, $\beta$ must be negative in order that $S$ be positive. This leads to concave $S$ and Eq.(23) is still a maximum. Boltzmann exponential distribution with negative temperature is an example.

As a matter fact, the only allowable minimum of entropy is zero in random dynamics. Zero entropy implies regular dynamics. Hence, in general, if entropy has a non zero extremum, it must be a maximum.

11) Nonequilibrium thermodynamics and maximum "nentropy"

  *a) Virtual work and the dynamical uncertainties*

In the above section, the virtual work was calculated for equilibrium system whose distribution $p_j$ does not depend explicitly on time.

Now we look at an ensemble of a large number of identical systems out of equilibrium. At a given moment of the motion, the systems are distributed over the time dependent microstates in the same way as an ensemble of equilibrium systems distributed over the time independent microstates. Suppose in this ensemble the systems composed of $N$ particles move from a initial point $a$ to a given point $b$ in the $3N$ dimensional configuration space. Without loss of generality, we can consider discrete paths denoted by $k=1,2$ ... (if the variation of the paths is continuous, the sum over $k$ must be replaced by path integral between $a$ and $b$[39]). At a given moment $t$ after the departure of the ensemble from the initial state, all the systems are distributed over different trajectories, each one arriving at a microstate $j$ ($j=1,2,$ ... w). Let $p_j$ be the probability that a system is found at the state $j$ having a value $x_j$ of a random variable $x$. The ensemble average of the variable at that moment is then given by $\bar{x} = \sum_{j=1}^{w} x_j p_j$.

A single system, after leaving its initial state, will be found at a given moment on the trajectory $k$ or at the corresponding microstate $j$. At this moment, the total force on a particle $i$ is $\vec{F}_i$ and its acceleration is $\vec{a}_i$ with an inertial force $-m_i \vec{a}_i$. Suppose a virtual displacement $\delta \vec{r}_i$ for the particle *i* at the state *j*, the virtual work over this displacement is given by $\delta W_{ij} = (\vec{F}_i - m_i \vec{a}_i)_j \cdot \delta \vec{r}_i$. Summing this work over all the particles, we obtain

$$\delta W_j = \sum_{i=1}^{N} (\vec{F}_i - m_i \vec{a}_i)_j \cdot \delta \vec{r}_i \quad (29)$$

To calculate the virtual work of $\vec{F}_i$, we will distinguish the forces of different nature. Let $\vec{f}_i$ be the interaction forces between the particles whose sum vanishes for the global system just as in the case of equilibrium system. These forces can be given by $\vec{f}_i = -\nabla w_i$ where $w_i$ is the potential energy. Let $\vec{\varphi}_{li}$ be the forces due to the gradient of certain variables such as pressure, temperature, particle density, chemical potentials etc (with $l=1,2,\ldots$). These forces can be written as $\vec{\varphi}_{li} = -\mu_l \nabla V_{li}$ where $V_{li}$ is the effective thermodynamic potential at the point of the particle *i* and $\mu_l$ is a constant characterizing the nature of $V_{li}$. Let $\vec{\phi}_i$ be an external force we can control by changing a parameter $\lambda$ such as volume, electric or magnetic field etc. Without considering the random virtual work at this stage, the virtual work of $\vec{F}_i$ then reads

$$\sum_{i=1}^{N} \vec{F}_i \cdot \delta \vec{r}_i = -\sum_i (\nabla w_i + \sum_l \mu_l \nabla V_{li} - \vec{\phi}_i)_j \cdot \delta \vec{r}_i = -\sum_{i=1}^{N} (\delta w_i + \sum_l \mu_l \delta V_{li} - \delta W_{\lambda i})_j. \quad (30)$$

where $\delta w_i = \nabla w_i \cdot \delta \vec{r}_i$ and $\delta V_{li} = \nabla V_{li} \cdot \delta \vec{r}_i$ are the virtual variation in the potentials due to the virtual displacement, $\delta W_{\lambda i} = \vec{\phi}_i \cdot \delta \vec{r}$ is the virtual work done to the particle *i* by changing the parameter $\lambda$.

For the second terms with inertial forces at the right hand side of Eq.(29), we can calculate $m \vec{a}_i \cdot \delta \vec{r}_i = m \delta \dot{\vec{r}}_i \cdot \dot{\vec{r}}_i = \delta(\frac{1}{2} m \dot{\vec{r}}_i^2) = \delta t_i$ where $t_i$ is the kinetic energy of the particle *i*. $\delta t_i$ can be still separated into two parts: the variation $\delta t_i^{eq}$ due to $\vec{f}_i$ and $\delta t_i^{neq}$ due to $\varphi_{il}$. It follows that

$$\delta W_j = -\sum_{i=1}^{N} (\delta w_i + \delta t_i)_j - \sum_{i=1}^{N} (\sum_l \mu_l \delta V_{li} - \delta W_{\lambda i})_j = -\sum_{i=1}^{N} (\delta e_i + \sum_l \mu_l \delta V_{li} + \sum_l \delta t_{il}^{neq} - \delta W_{\lambda i})_j. \quad (31)$$

where $\delta e_i$ is the virtual variation of the equilibrium part (i.e., $e_i = t_i^{eq} + w_i$) of the total energy of the particle for the microstate *j* (remember that $w_i$ is defined as the potential of the equilibrium forces). In this way, the virtual work $\delta W_j$ is separated into two parts, i.e., $\delta W_j = \delta W_j^{eq} + \delta W_j^{neq}$. The first is the virtual work for equilibrium state $\delta W_j^{eq} = -\sum_{i=1}^{N}(\delta e_i)_j$ and the second is for nonequilibrium states $\delta W_j^{neq} = -\sum_{i=1}^{N}(\sum_l \mu_l \delta V_{li} + \delta t_{il}^{neq} - \delta W_{\lambda i})_j$.

The equilibrium virtual work being investigated in the last section, below we will see into the *nonequilibrium* virtual work $\delta W_j^{neq} = -\sum_{i=1}^{N}(\sum_l \mu_l \delta V_{li} + \delta t_{il}^{neq} - \delta W_{\lambda i})_j = -\sum_{i=1}^{N}(\delta \varepsilon_i - \delta W_{\lambda i})_j$ where $\varepsilon_i = \sum_l (\mu_l V_{li} + t_{il}^{neq})$ can be referred to as the nonequilibrium 'energy' of a particle (not necessarily an energy in the usual sense since $V_{li}$ is not the usual potential energy). This virtual work can also be expressed over one particle states $k=1,2 \ldots g$:

$$\delta W_j^{neq} = -\sum_{k=1}^{g}(n_k)_j \delta \varepsilon_k + \delta W_\lambda = -\delta \sum_{k=1}^{g}(n_k)_j \varepsilon_k + \sum_{k=1}^{g} \varepsilon_k (\delta n_k)_j + \delta W_\lambda \qquad (32)$$
$$= -\delta \mathrm{E}_j + \sum_{k=1}^{g} \varepsilon_k (\delta n_k)_j + \delta W_\lambda$$

where $\mathrm{E}_j = \sum_k (n_k)_j \varepsilon_k$ is the total nonequilibrium energy of the system and $\delta W_{\lambda j} = \sum_i (\delta W_{\lambda i})_j$ the total external work at the state $j$ due to the change of the parameter $\lambda$. The average nonequilibrium virtual work is given by

$$\overline{\delta W}^{neq} = \sum_{j=1} p_j \delta W_j^{neq} = -\overline{\delta \mathrm{E}} + \sum_{k=1}^{g} \varepsilon_k \overline{\delta n_k} + \overline{\delta W_\lambda}. \qquad (33)$$

where $\overline{\delta \mathrm{E}} = \sum_j p_j \delta \mathrm{E}_j$ is the average of energy variation, the second term $\sum_{k=1}^{g} \varepsilon_k \overline{\delta n_k}$ is the variation of the nonequilibrium energy due to $\overline{\delta N}$, the average variation of the particle number and the last term $\overline{\delta W_\lambda} = \sum_j p_j \delta W_{\lambda j}$ is the average external virtual work. We write $\sum_k \varepsilon_k \overline{\delta n_k} = \omega \sum_j p_j \delta N_j = \omega \overline{\delta N}$ with $\omega = \sum_k \varepsilon_k \overline{\delta n_k} / \overline{\delta N}$ which is a quantity imitating the equilibrium chemical potential. It represents the change of nonequilibrium energy of the system due to the change of particle number. It follows that

$$\overline{\delta W}^{neq} = -\overline{\delta \mathrm{E}} + \omega \overline{\delta N} + \overline{\delta W_\lambda} = -\delta \overline{\mathrm{E}} + \omega \delta \overline{N} + \overline{\delta W_\lambda} + \sum_j (\mathrm{E}_j - \omega N_j - W_{\lambda j}) \delta p_j \qquad (34)$$

with $\bar{E} = \sum_j p_j E_j = \sum_j p_j \sum_k (n_k)_j \varepsilon_k$ is the nonequilibrium energy, $\overline{W_\lambda} = \sum_j p_j W_{\lambda j}$ is the cumulate external work performed by changing $\lambda$ from $t_0$ the beginning of the nonequilibrium process up to the moment $t$ of the virtual displacement.

What is the quantity $\sum_j (E_j - \omega N_j - W_{\lambda j}) \delta p_j$? Due to the presence of $\bar{E}$ (not the usual energy) and of the work $W_{\lambda j}$ in its expression, we are quite sure that this uncertainty is not the thermodynamic entropy $\delta S^{eq} = \beta \sum_j (E_j - \mu N_j) \delta p_j$. Let it be called "nentropy" and denoted by

$$\delta S^{neq} = \eta \sum_j (E_j - \omega N_j - W_{\lambda j}) \delta p_j \tag{35}$$

where $\eta$ is a characteristic parameter of the nonequilibrium state.

The total virtual work for the nonequilibrium system reads

$$\overline{\delta W} = -\delta \bar{E} - \delta \bar{E} + (\mu + \omega) \delta \bar{N} + \delta \overline{W_\lambda} + \delta S^{neq}/\eta + \delta S^{eq}/\beta \tag{36}$$

Let $\delta \Omega = \delta S^{neq}/\eta + \delta S^{eq}/\beta$ represent the total uncertainty of the dynamics, applying the virtual work principle $\overline{\delta W} = 0$ to this equation yields

$$\delta \Omega - \delta \bar{E} - \delta \bar{E} + (\mu + \omega) \delta \bar{N} + \delta \overline{W_\lambda} = 0 \tag{37}$$

or $\delta(S^{eq}/\beta + S^{neq}/\eta - \bar{E} - \bar{E} + (\mu + \omega) \bar{N} + \overline{W_\lambda}) = 0$ for the nonequilibrium dynamics.

*b) Maximum nentropy*

In what follows, for the sake of simplicity, we only discuss the case of closed system without variation of particle number, i.e., $\delta \bar{N} = 0$ (canonical ensemble). Eq.(37) now reads:

$$\delta(S^{eq}/\beta + S^{neq}/\eta - \bar{E} - \bar{E} + \overline{W_\lambda}) = 0. \tag{38}$$

The dividing of nonequilibrium dynamics into an equilibrium part and a nonequilibrium part makes it possible to tackle the two different uncertainties $\delta S^{eq}$ and $\delta S^{neq}$ separately. This approach will help us to understand more about the variational recipe of Eq.(38), especially the nonequilibrium component in it.

We already proved the $\delta W^{eq} = \delta(S^{eq} - \beta \overline{E}) = 0$ for equilibrium system. This equilibrium part of virtual work is in general independent from the nonequilibrium part. Hence $\delta W^{eq} = 0$ still holds in nonequilibrium dynamics. This leads to $\delta W^{neq} = 0$ and

$$\delta[S^{neq} - \eta(\overline{E} - \overline{W_\lambda})] = 0. \tag{39}$$

This extremum of $S^{neq}$ is different from maximum thermodynamic entropy for several reasons. The first reason is that $S^{neq}$ is not thermodynamic entropy as mentioned above. This can be seen from the difference between its definition Eq.(35) and the definition of the thermodynamic entropy, Eq.(21) (the mathematical form $\delta S^x = \eta \sum_j x_j \delta p_j$ is referred to as varentropy in the references [26][27] in order to distinguish it from entropy). The second reason is that, due to Eq.(39), it is function of an unusual energy $\overline{E}$ and the cumulate external work $\overline{W_\lambda}$ contributing to the nonequilibrium process, contrary to the equilibrium case where $S^{eq}$ is function of only the real energy $\overline{E}$. The third reason is that the thermodynamic entropy $S^{eq}$ can be proved to be concave with respect to the equilibrium probability variation, but the same reasoning cannot be made for $S^{neq}$ since in general the nonequilibrium state distribution $p_j$ is not known. Since $\overline{E}$ and $\overline{W_\lambda}$ are the energy associated with macroscopic transfer such as matter diffusion, heat conduction, macroscopic expansion of gas etc, $S^{neq}$ should be the uncertainty of the transfer processes and remains constant or vanishing without transport phenomenon (equilibrium). Hence $\delta S^{neq}$ is the production of the total uncertainty $\Omega = S^{neq}/\eta + S^{eq}/\beta$ by the transfer phenomena. Eq.(39) is the extremum algorithm of that production and can be called extremum nentropy production (ENP).

One should notice that ENP deduced from the virtual work principle is not necessarily the maximum entropy production (MEP)[9] associated with the thermodynamic entropy of the second law, since $S^{neq}$ is not $S^{eq}$ in general. ENP needs not the hypothesis of local equilibrium one must consider in order to use thermodynamic entropy. Consequently, it can be used for any nonequilibrium process close to or far from equilibrium for which the thermodynamic entropy is not defined.

Eventually ENP can be used for assigning nonequilibrium probability distribution over the instantaneous microstates if ever $S^{neq}$ is known as a functional of nonequilibrium probability distribution, i.e., $S^{neq} = f(p_1, p_2, ... p_w)$. We can write $\delta S^{neq} = \sum_j \frac{\partial f}{\partial p_j} \delta p_j$ as a result of a virtual variation in the probability. On the other hand, we have $\delta S^{neq} = \eta \sum_j (E_j + W_{\lambda j}) \delta p_j$ through Eq.(35) which implies

$$\sum_j (\frac{\partial f}{\partial p_j} - \eta E_j + \eta W_{\lambda j}) \delta p_j = 0. \qquad (40)$$

By virtue of the normalization condition $\sum_{j=1}^{w} \delta p_j = 0$ for this given moment, one can prove [28] that

$$\frac{\partial f}{\partial p_j} - \eta (E_j - W_{\lambda j}) = K. \qquad (41)$$

with a constant $K$. Eq.(41) can be used for deriving the probability distribution. Inversely, if the probability distribution is known, one can derive the functional of $S^{neq}$ through $\delta S^{neq} = \eta \sum_j (E_j - W_{\lambda j}) \delta p_j$ [26][27].

### c) Application to free expansion of ideal gas

A question arises about the difference between $S^{neq}$ and $S^{eq}$ especially when two equilibrium states are related by a nonequilibrium process. In what follows, we discuss some cases where the variation $\Delta S^{neq}$ during the nonequilibrium process is just the difference of the second law entropy, i.e., $S^{neq} = \Delta S^{eq} = S_2^{eq} - S_1^{eq}$ between the two equilibrium states.

We consider only one effective potential $V_l$ in the case of the diffusion due to density gradient. Let $C$ be the density of the diffused particles and $V_l = \ln C$ be the effective thermodynamic potential[38][11]. The thermodynamic force is $\bar{\varphi}_l = -\nabla V_l = -\frac{\nabla C}{C}$ ($\mu_l = 1$). This leads to $J = Cu = C\mu_d \varphi_l$ ($u = \mu_d \varphi_l$ is the drift velocity of a particle with mobility $\mu_d$) and to the Fick's first law $J = -D\nabla C$ with the diffusion constant $D = \mu_d$.

Eq.(39) is an equality expressing a virtual variation relationship at a given position. Yet it can be used to calculate the true nentropy production in a transport process. The calculation is shown below.

By definition, $\overline{E} = \sum_j p_j \sum_{i=1}^N (V_{li} + t_{il}^{neq})_j = \sum_j p_j \sum_{i=1}^N (V_{li})_j + \sum_j p_j \sum_{i=1}^N (t_{il}^{neq})_j = V_l + T_l$ where $T_l = \sum_j p_j \sum_{i=1}^N (t_{il}^{neq})_j$ is the total kinetic energy of the particles associated with the nonequilibrium process, and $V_l = \sum_j p_j \sum_{i=1}^N (V_{li})_j$ is the average total potential energy of the system that was given above by $V_l = \ln C$. Eq.(39) now reads:

$$\delta S^{neq} = \eta(\delta V_l + \delta T_l - \delta \overline{W_\lambda}) = \eta \delta \ln C + \eta \delta T_l - \eta \delta \overline{W_\lambda}. \tag{42}$$

This result can be applied to the free adiabatic expansion of ideal gas of $N$ particles. Since the total system is isolated, the total kinetic energy is constant ($\delta T_l = 0$) and there is no external work, we simply have $\delta S^{neq} = \eta \delta \ln C$.

Now considering an expansion from state 1 to state 2, both in equilibrium with respectively volumes $V_1$ to $V_2$ and uniform densities $C_1$ and $C_2$, we straightforwardly write

$$\Delta S^{neq} = \eta(\ln C_2 - \ln C_1) = -\eta \ln \frac{V_2}{V_1} \tag{43}$$

which is just the well known result $\Delta S^{eq} = S_2 - S_1 = Nk_B \ln \frac{V_2}{V_1}$ provided $\eta = -Nk_B$ where $k_B$ is the Boltzmann constant and $N$ the total number of diffused particles. As mentioned above, this is a case where the nentropy production is just the entropy production of a process (far from equilibrium) between two equilibrium states.

When a steady state exists, the time independent gradient $\nabla C$ over the whole system or over a small volume $\delta V$ of the system makes sense. We can write $\delta S^{neq} = \eta \delta \ln C = \eta \nabla \ln C \cdot \delta \vec{r}$ in that volume where $\delta \vec{r}$ is an elementary displacement in the direction of the transport flux $\vec{J} = n\vec{u}$ where $n$ is the density of particles in that volume and $\vec{u}$ the drift velocity. Let $\delta \vec{r} = \vec{u} \delta t$ in the time interval $\delta t$, the nentropy reads

$$\Delta S^{neq} = -n \delta V k_B \nabla \ln C \cdot \vec{u} \delta t. \tag{44}$$

This yields the nentropy production rate given by

$$\sigma^{neq} = \frac{\Delta S^{neq}}{\delta V \delta t} = -k_B \nabla \ln C \cdot \vec{J} = \vec{F} \cdot \vec{J} \tag{45}$$

where $F = -k_B \nabla \ln C$ is a thermodynamic force. This is the same expression as the entropy production rate for steady state of nonequilibrium process[11].

*d) Application to heat conduction in solid*

Within some approximation, we can consider the heat flow as a phonon gas expansion due to temperature difference in an *isolated body*. Here the energy is conserved, hence the total number of phonons can be considered constant.

We consider an insulating solid in which a heat flow is created by temperature difference. Suppose that a vibrating atom contains $n_\nu$ phonons in a mode of frequency $\nu$ and that the expectation of $n_\nu$ is given by the Bose-Einstein distribution

$$n_\nu = \frac{1}{\exp(\frac{h\nu}{k_B T}) - 1} \tag{46}$$

where $h$ is the Planck constant and the chemical potential $\omega$ is taken to be zero. The phonon concentration of the modes in the interval $\nu \to \nu + d\nu$ is then $C_\nu = g(\nu) n_\nu d\nu$ where $g(\nu)$ is the number of modes per unity of volume in that interval. The thermodynamic force for this mode is $\bar{\varphi}_{l,\nu} = -\nabla C_\nu / C_\nu = \chi_\nu \nabla\left(\frac{1}{T}\right)$. The flux of phonons of this mode is given by

$$\bar{I}_\nu = -C_\nu \mu_{d,\nu} \frac{\nabla C_\nu}{C_\nu} = C_\nu \mu_{d,\nu} \chi_\nu \nabla\left(\frac{1}{T}\right) \quad \text{where} \quad \chi_\nu = \frac{\exp(h\nu / k_B T)(h\nu / k_B)}{\exp(h\nu / k_B T) - 1}$$

and $\mu_{d,\nu}$ is the mobility of phonons of the mode $\nu$. The heat flux for these modes is given by $\bar{J}_\nu = I_\nu h\nu$. The transport equation of heat conduction can be found by integrating $J_\nu$ over all the modes, i.e.[11],

$$\bar{J} = L \nabla\left(\frac{1}{T}\right) \tag{47}$$

where the phenomenological transport coefficient is given by $L = \int_0^{\nu_D} n_\nu h\nu \mu_{d,\nu} \chi_\nu g(\nu) d\nu$ with $\nu_D$ the Debye frequency of the crystal. The Fourier law is $\bar{J} = -\kappa \nabla T$ with the heat conductivity $\kappa = \frac{L}{T^2}$. Notice that $\kappa$ and $L$ may depend on temperature.

When temperature is so high that $\frac{h\nu}{k_B T} \ll 1$, we have $C_\nu = \frac{g(\nu)k_B T d\nu}{h\nu}$, $\chi_\nu = T$ and $\vec{J}_\nu = k_B T^2 \mu_{d,\nu} g(\nu) d\nu \nabla\left(\frac{1}{T}\right)$. So and $\vec{J} = -\kappa \nabla T$ where $\kappa = k_B \int_0^{\nu_D} \mu_{d,\nu} g(\nu) d\nu$.

We can consider a process between two states where the temperatures are well defined (two equilibrium or local equilibrium states for example). From Eq.(42) with $\delta T_l = 0$ and $\delta \overline{W}_\lambda = 0$ (isolated system), choosing $V_l = -\frac{1}{T}$ for the two states, we can write

$$\delta S^{neq} = \eta \mu_l \delta V_l = \eta \mu_l \delta\left(-\frac{1}{T}\right) = \alpha \delta \ln T \tag{48}$$

where $\alpha = \frac{\eta \mu_l}{T} = \eta \int_0^{\nu_D} g(\nu) d\nu$ is a constant.

Now let us look into the case of entropy increase in an isolated system containing two solids which were initially in equilibrium at two different temperatures $T_1$ and $T_2$ ($T_1 > T_2$), and are then equilibrated at a common temperature $T_f$ by thermal contact. Due to the heat flow from solid 1 to solid 2, the temperature of solid 1 (or 2) decreases (or increases) and gradually approaches $T_f$. The nentropy production per unit volume in solid 1 is $\Delta S_1^{neq} = \alpha \ln \frac{T_f}{T_1}$ and in solid 2 is $\Delta S_2^{neq} = \alpha \ln \frac{T_f}{T_2}$. The total nentropy production per unit volume is thus given by

$$\Delta S^{neq} = \Delta S_1^{neq} + \Delta S_2^{neq} = \alpha \ln \frac{T_f}{T_1} + \alpha \ln \frac{T_f}{T_2} = \alpha \ln \frac{T_f^2}{T_1 T_2}.$$ If we let $\alpha = c_v$, the specific heat of the system, $\Delta S^{neq}$ is then the entropy production $\Delta S^{eq} = c_v \ln \frac{T_f^2}{T_1 T_2}$ of the process of equilibration by thermal contact. We stress again that $\Delta S^{neq}$ is the entropy production if and only if the process takes place between two equilibrium states.

For a steady state, a potential $V_l = -\frac{1}{T}$ and $\vec{\varphi}_l = -\mu_l \nabla\left(-\frac{1}{T}\right)$ can be defined everywhere in the system[11] with $\mu_l = \int_0^{\nu_D} \chi_\nu g(\nu) d\nu = T \int_0^{\nu_D} g(\nu) d\nu$. Similar expression to Eqs.(43-44) for heat conduction can be found, i.e.,

$$\sigma^{neq} = \frac{\Delta S^{neq}}{\delta t} = \vec{F} \cdot \vec{J} \qquad (49)$$

where $\vec{F} = -\nabla(-\frac{1}{T})$ is the thermodynamic force and $\vec{J} = C_{heat}\vec{u}$ is the heat flux with $C_{heat} = k_B T \int_0^{\nu_D} g(\nu)d\nu$ can be interpreted as the heat density in the system.

## 12) Reversible process

A system can undergo variation of states through very slow transition from states to states, such as in the usual quasi-equilibrium process treated in the equilibrium thermodynamics with the first law and the second law with equality. Eq.(38) should be valid in this case as well. Since there is no transport, we have $\delta S^{neq} = 0$ and $\delta \overline{E} = 0$. Eq.(38) becomes

$$\delta(S^{eq}/\beta - \overline{E} + \overline{W_\lambda}) = 0. \qquad (50)$$

Or

$$\delta \overline{E} = \delta S^{eq}/\beta + \delta \overline{W_\lambda} \qquad (51)$$

which is the usual combination of the first and second laws in equilibrium thermodynamics.

## 13) Stochastic least action principle (SAP)

The least action principle[30][33] was formulated for regular dynamics of mechanical system. One naturally ask the question about its fate when the system is subject to noise making the dynamics irregular. In order to answer this question, a stochastic action principle

$$\overline{\delta A} = 0 \qquad (52)$$

has been developed in [21] where $\delta A$ is a variation of the Lagrange action $A$ and the average is carried out over all possible paths between two given points in configuration space. Eq.(52) was postulated as a hypothesis in the previous work. Here we will give a derivation from the virtual work principle for random dynamics.

We consider a statistical ensemble of mechanical systems out of equilibrium and its trajectories in configuration space. Each system is composed of $N$ particles moving in the 3$N$ dimensional space starting from a point $a$. If the motion was regular, all the systems in the ensemble would follow a single 3$N$-dimensional trajectory from $a$ to a given point $b$ according to the least action principle. In random dynamics, every system can take different paths from *a* to *b* as discussed in previous sections.

Now let us look at the random dynamics of a single system following a trajectory, say, *k*, from *a* to *b*. At a given time *t*, the total force on a particle *i* in the system is denoted by $\vec{F}_i$ and the acceleration by $\vec{a}_i$ with an inertial force $-m_i\vec{a}_i$ where $m_i$ is its mass. The virtual work at this moment on a virtual displacement $\delta \tilde{r}_{ik}$ of the particle *i* on the trajectory *k* reads

$$\delta W_i = [\vec{F}_i - m_i \vec{a}_i]_k \cdot \delta \tilde{r}_{ik} \tag{53}$$

Summing this work over all the particles, we obtain

$$\delta W_k = \sum_{i=1}^{N} (\vec{F}_i - m_i \vec{a}_i)_k \cdot \delta \tilde{r}_{ik} \tag{54}$$

This virtual work is similar to $\delta W_j$ of Eq.(31) since a trajectory leads necessarily to a microstate at a given moment of time. The average virtual work previously given by Eq.(13) can also be calculated by $\overline{\delta W} = \sum_{k=1}^{w} p_k \delta W_k$ where we considered discrete paths denoted by *k*=1,2 … *w* (if the variation of path is continuous, the sum over *k* must be replaced by path integral between *a* and *b*[39]), and $p_k$ is the probability that the path *k* is taken.

Now let us establish the relationship between virtual work and action variation (for one dimensional case, *x* is the position in the configuration space). For a given path *k*, the action variation is given by

$$\delta A_k = \sum_{i=1}^{N} \int_a^b \delta L_{ik} dt = \sum_{i=1}^{N} \int_a^b \left( \frac{\partial L}{\partial x_i} \delta x_i + \left( \frac{\partial L}{\partial \dot{x}} \right) \delta \dot{x}_i \right)_k dt = \int_a^b \sum_{i=1}^{N} \left( -\frac{\partial H_i}{\partial x_i} - \dot{P}_{xi} \right)_k \delta x_{ik} dt \tag{55}$$

$$= \int_a^b \sum_{i=1}^{N} (F_{xi} - m\ddot{x}_i)_k \delta x_{ik} dt = \int_a^b \delta W_k dt$$

where we used, for the particle *i* with Hamiltonian $H_i$ and Lagrangian $L_i$, $F_{xi} = -\frac{\partial H_i}{\partial x_i} = \frac{\partial L_i}{\partial x_i}$, $m_i \ddot{x}_i = \dot{P}_{xi} = \frac{\partial}{\partial t}\left(\frac{\partial L_i}{\partial \dot{x}_i}\right)$ and $\int_a^b \frac{\partial}{\partial t}\left(\delta x_j \frac{\partial L}{\partial \dot{x}}\right) dt = \left(\delta x_j \frac{\partial L}{\partial \dot{x}}\right)_a^b = 0$ because of the zero variation at *a* and *b*.

The average action variation being $\overline{\delta A} = \sum_{k=1}^{w} p_k \delta A_k = \int_a^b \overline{\delta W} dt$, the virtual work principle Eq.(13) yields Eq.(52), i.e., $\overline{\delta A} = 0$. This SAP implies an varentropy variational approach. To see this, we calculate

$$\overline{\delta A} = \delta \sum_{j=1}^{w} p_k A_k - \sum_{j=1}^{w} \delta p_k A_k = \delta \overline{A}_{ab} - \delta Q_{ab} \qquad (56)$$

where $\overline{A}_{ab} = \sum_{k=1}^{w} p_k A_k$ is the ensemble mean of action $A_k$ between $a$ and $b$, and $\delta Q_{ab}$ can be written as

$$\delta Q_{ab} = \delta \overline{A}_{ab} - \overline{\delta A} = \sum_{j=1}^{w} A_k \, \delta p_k . \qquad (57)$$

which is a varentropy measuring the uncertainty in the choice of trajectories by the system. We can introduce a path entropy $S_{ab}$ such that

$$\delta Q_{ab} = \frac{\delta S_{ab}}{\gamma} . \qquad (58)$$

Then Eq.(52) and Eq.(56) yield

$$\delta(S_{ab} - \gamma \overline{A}_{ab}) = 0 . \qquad (59)$$

If the normalization condition is added as a constraint of variational calculus, Eq.(59) becomes

$$\delta [S_{ab} - \gamma \sum_k p_k A_k + \alpha \sum_k p_k] = 0 \qquad (60)$$

This is a maximum path entropy with two Lagrange multipliers $\alpha$ and $\gamma$, an approach originally proposed in the references [21].

## 14) A path probability distribution

### a) Directed schema

It was shown[21] that, if the path entropy takes the Shannon form, i.e., $S_{ab} = -\sum_k p_k \ln p_k$, the SAP or Eq.(60) yields an exponential probability distribution of action

$$p_k(a,b) = \frac{1}{Z_{ab}} e^{-\gamma A_k(a,b)} . \qquad (61)$$

where $Z_{ab} = \sum_{k} e^{-\gamma A_k(a,b)}$, meaning that this distribution describes a motion directed from a fixed $a$ to a fixed $b$ (see fig.4).

The path entropy can be calculated by

$$S_{ab} = \ln Z_{ab} + \gamma \overline{A}_{ab}. \tag{62}$$

where $\overline{A}_{ab} = \sum_{k=1}^{w} p_k(a,b) A_k = -\dfrac{\partial}{\partial \gamma} \ln Z_{ab}$ is the average action between these two fixed points.

### b) Panoramic schema

The above description is not complete for the dynamics since a real motion from an initial point $a$ does not necessarily arrive at $b$. The system moves around and can reach any point in the final volume, say, $B$. Hence a complete description of the dynamics requires unfixed $b$ in $B$. The probability $p_k(a, B)$ for the system to go from a fixed point $a$ to a unfixed $b$ through a certain path $k$ (depending on $a$ and $b$) is given by

$$p_k(a, B) = \frac{1}{Z_a} e^{-\gamma A_k(a,b)}. \tag{63}$$

where $Z_a = \sum_{b,k} e^{-\eta A_k(a,b)} = \sum_b Z_{ab}$. Hence the path entropy for this case is given by

$$S_{aB} = \ln Z_a + \gamma \overline{A}_a. \tag{64}$$

Here $\overline{A}_a = \sum_b \sum_k p_k(a,B) A_k(a,b) = -\dfrac{\partial}{\partial \gamma} \ln Z_a$ is the average action over all the paths between a fixed point $a$ to all the points in the final volume B. We have the following relationship

$$\overline{A}_a = \sum_b \sum_k p_k(a,B) A_k(a,b) = \sum_b \frac{Z_{ab}}{Z_a} \overline{A}_{ab} = \sum_b \frac{\exp(\ln Z_{ab})}{Z_a} \overline{A}_{ab} = \sum_b \frac{\exp(S_{ab} - \gamma \overline{A}_{ab})}{Z_a} \overline{A}_{ab}$$

The function $p(a, B) = \dfrac{\exp(S_{ab} - \gamma \overline{A}_{ab})}{Z_a}$ is the probability from the point $a$ to an arbitrary point $b$ in the final volume B no matter what path the process may take.

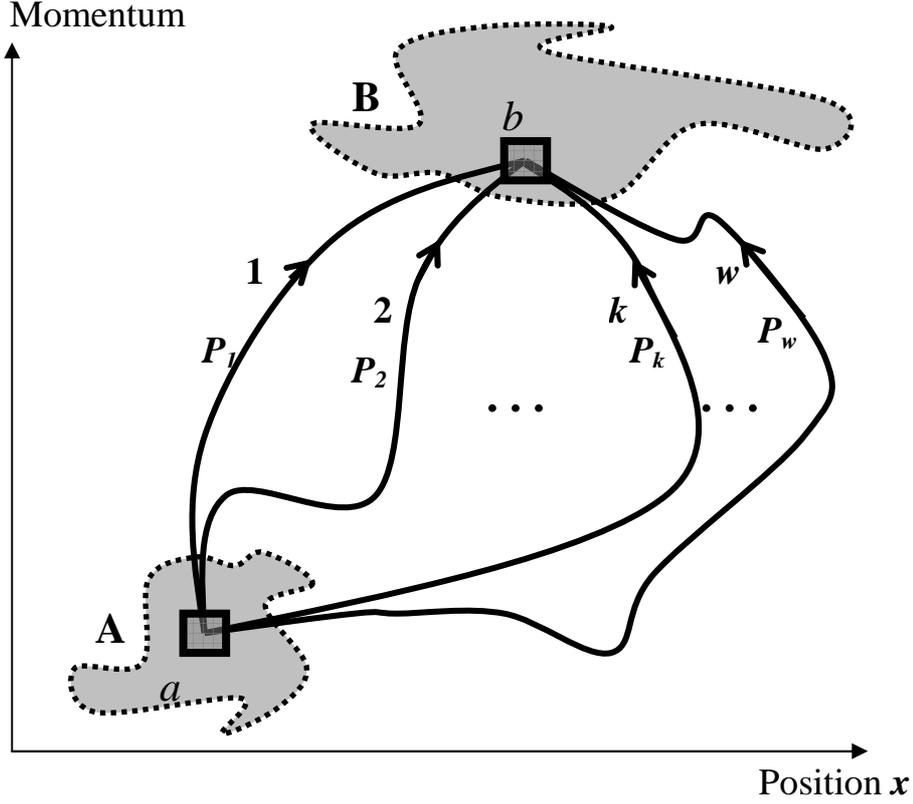

**Figure 4**: Illustration of the 3 schemas of a random dynamics in phase space. **A** is the initial volume at time $t_a$ and $a$ is any point in **A**. **B** is the final volume at time $t_b$ and $b$ is any point in **A**. The *directed schema* means the system, leaving from a certain $a$, must arrive at a fixed $b$ in **B**. The *panoramic schema* means the system, leaving from certain $a$, arrives at any arbitrary point $b$ in **B**. The *initial condition schema* adds the uncertainty in the initial condition, meaning that the system, arriving at certain point $b$ in **B**, can come from any arbitrary point $a$ in **A**.

*c) Initial condition schema*

In order to include the contribution of the initial conditions to the dynamic uncertainty, we extend still the path probability to the schema in which $a$ is also relaxed in the initial volume $A$. The transition probability from $A$ to $B$ through a certain path $k$ is

$$p_k(A,B) = \frac{1}{Z} e^{-\gamma A_k(a,b)}. \tag{65}$$

where $Z = \sum_{a,b,k} e^{-\gamma A_k(a,b)} = \sum_a Z_a$. The total path entropy between A and B reads

$$S_{AB} = \ln Z + \gamma \overline{A}. \tag{66}$$

Here $\overline{A} = \sum_a \sum_b \sum_k p_k(A,B) A_k(a,b) = -\frac{\partial}{\partial \gamma} \ln Z$ is the average action of the process from A to B.

The total transition probability $p(A,B)$ between an arbitrary point $a$ in A to an arbitrary point $b$ in B through whatever paths is given by

$$p(A,B) = \frac{1}{Z} \exp(S_{ab} - \gamma \overline{A}_{ab}). \tag{67}$$

Using a Legendre transformation $F_{ab} = \overline{A}_{ab} - S_{ab}/\gamma = \frac{1}{\gamma} \ln Z_{ab}$ which can be called *free action* mimicking the free energy of thermodynamics, we can write $p(A,B) = \frac{1}{Z} \exp(-\gamma F_{ab})$.

This section provides a series of path probability distributions in exponential of action describing the likelihood of each path to be chosen by the motion. It is clear that if the constant $\gamma$ is positive, the most probable path will be least action path. This implies that if the randomness of motion is vanishing, all the paths will collapse onto the bundle of least action ones, which is accordance with the least action principle of regular motion. A more mathematical discussion can be found in [21]. This formalism is to some extent a classical version of the idea of M. Gell-Mann[32] to characterize, in superstring theory, the likelihoods of different solutions of the fundamental equation by quantized and Euclideanized action.

15) SAP and Brownian motion

*a) Path probability*

A possible application of SAP is the Brownian particles. This has been discussed in some detail in our previous work[21][22][23][24]. The exponential distribution of action can be shown with the help of the transition probability of Brownian motion which is given by $p_{ij} = \frac{1}{Z_{ij}} \exp\left[-\frac{(x_j - x_i)^2}{4D(t_j - t_i)}\right]$ from a position $x_i$ to another one $x_j$ where $Z_{ij} = \sqrt{4\pi D(t_j - t_i)}$ for one dimensional motion[21]. This means that, the probability of a path $k$ composed of a series of very large $N$ successive and very close positions { $x_a$, $x_1$, $x_2$ … $x_{N-1}$, $x_b$ } is

$$p_k(a,b) = p_{a1} p_{12} p_{23} \ldots p_{N-1,b} = \frac{1}{Z} \exp\left[-\frac{(x_1 - x_a)^2}{4D(t_1 - t_a)} - \frac{(x_2 - x_1)^2}{4D(t_2 - t_1)} \ldots - \frac{(x_b - x_{N-1})^2}{4D(t_b - t_{N-1})}\right] \tag{68}$$

If the successive positions are infinitesimally close to each other, the exponent
$\frac{m(x_1-x_a)^2}{2(t_1-t_a)^2}(t_1-t_a)+\frac{m(x_2-x_1)^2}{2(t_2-t_1)^2}(t_2-t_1)...+\frac{m(x_b-x_{N-1})^2}{2(t_b-t_{N-1})^2}(t_b-t_{N-1})$ is just the action $A_k(a,b)$ of the particle of mass $m$ from $a$ to $b$ along the path $k$. Hence $p_k(a,b)=\frac{1}{Z}e^{-\gamma A_k(a,b)}$ with $\frac{1}{Z}=C\frac{1}{Z_{a1}}\frac{1}{Z_{12}}...\frac{1}{Z_{N-1,b}}$, $\gamma=\frac{1}{2mD}$ and $C$ is a constant depending on which of the above three schemas is used (see below).

This analysis is confirmed by numerical simulations[47]. However, since the potential energy is zero in the present case, there is no difference between action and the time average of Hamiltonian $\overline{H}_k=\frac{1}{t_b-t_a}\int_a^b\frac{1}{2}mv_k^2 dt=\frac{A_k}{t_b-t_a}$ where $v_k$ is the velocity of the motion along the path $k$. Some authors[40] believe that the path probability must be exponential of average Hamiltonian. A simple way to solve this ambiguity by numerical simulation is to add potential energy to the Brownian particle.

In the absence of such work to date, the following reasoning a priori argues in favor of the exponential of action. If the probability was exponential of average energy, the most probable path would be the lowest energy path. When the noise of the random dynamics becomes weaker and weaker, the dynamics will become less and less irregular and progressively tends to regular limit if finally the noise vanishes. In this case, the unique remaining path would be the least energy path. This is in conflict with least action principle yielding least action path for regular dynamics.

*b) Path integral of Brownian motion*

We give here an example of the path integral technique used for Brownian motion.

In order to calculate the partition function of the distribution in Eq.(68), the summation $\sum_k$ between two fixed points $a$ and $b$ should be replaced by $\int\omega^{N-1}dx_1dx_2...dx_{N-1}$ where $\omega$ is the uniform density of path in the configuration space. From the classical point of view, the paths are supposed continuous, i.e., there is no unoccupied or inaccessible space between the paths. This causes problem for the above replacement of $\sum_k$ since $\omega$ is infinite.

In what follows, we will introduce a method of discretization of path using the uncertainty relation $\overline{\Delta x^2}\cdot\overline{\Delta P^2} \geq \frac{1}{4\gamma^2}$ of the dynamics (see section 19) leading to the discretization of phase space by $\frac{dxdP}{1/\gamma}$. In order to discretize only the configuration space, we will, faute de mieux, consider $dP = m\frac{dx}{dt}$ and $\frac{mdx^2}{dt} \geq 1/\gamma$, meaning that, statistically, the smallest meaningful interval of position is $dx_{\min} = \sqrt{\frac{dt}{m\gamma}}$. Hence we can use $\omega = \sqrt{\frac{m\gamma}{2\pi dt}}$ ($2\pi$ is added for simplicity in the result, see below) as path density in the intervals from $x$ to $x+dx$ and from $t$ to $t+\tau$. This is equivalent to the definition of path integral in [48]. Fig.5 illustrates the path integral calculus in configuration space.

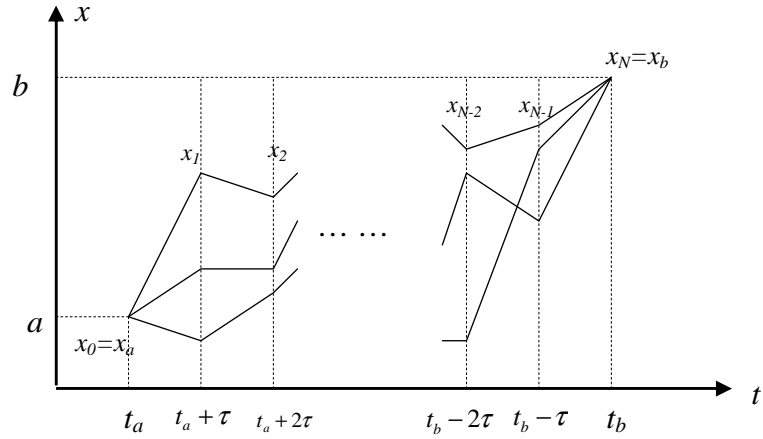

Figure 5: A geometrical illustration of the path integral between *a* and *b* in a 1D configuration space.

In the directed schema with $p_k(a,b) = \frac{1}{Z_{ab}}e^{-\gamma A_k(a,b)}$, the partition function from Eq.(68) is given by Gaussian integral:

$$Z_{ab} = \left(\frac{m\gamma}{2\pi\tau}\right)^{(N-1)/2} \int dx_1 dx_2 ... dx_{N-1} \exp\left[-\frac{(x_1-x_a)^2}{4D\tau} - \frac{(x_2-x_1)^2}{4D\tau} ... - \frac{(x_b-x_{N-1})^2}{4D\tau}\right] \quad (69)$$

$$= \sqrt{\frac{1}{N}}\exp\left(-\gamma\frac{m(x_b-x_a)^2}{2(t_b-t_a)}\right).$$

Then $\bar{A}_{ab} = -\frac{\partial}{\partial \gamma} \ln Z_{ab}$ yields

$$\bar{A}_{ab} = \frac{m(x_b - x_a)^2}{2(t_b - t_a)}. \tag{70}$$

For panoramic schema, we have

$$Z_a = \sum_b Z_{ab} = \left(\frac{m\gamma}{2\pi(t_b - t_a)}\right)^{1/2} \int_{-\infty}^{\infty} dx_b \exp\left(-\gamma \frac{m(x_b - x_a)^2}{2(t_b - t_a)}\right) = 1 \tag{71}$$

where we used $(t_b - t_a) = N\tau$. From $\bar{A}_a = -\frac{\partial}{\partial \gamma} \ln Z_a$, one obtains $\bar{A}_a = 0$. These results are logical for free Brownian particles.

## 16) Hamiltonian mechanics revisited

It has been shown in our previous work[23] that the exponential probability distribution of action satisfied the Fokker-Planck equation for normal diffusion in the same way as the Feymann factor of quantum propagator $P_k \propto e^{-iA_k/\hbar}$ satisfies the Schrödinger equation[39]. Here we will show a modified formalism of Hamiltonian mechanics within the present approach.

*a) Euler-Lagrange equations*

A meaning of the SAP Eq.(52) is that for any particular path $k$, we do not have necessarily $\delta A_k = 0$. Hence we have

$$\delta A_k = \int_a^b \left[\frac{\partial}{\partial t}\left(\frac{\partial L_k}{\partial \dot{x}}\right) - \frac{\partial L_k}{\partial x}\right] \delta x \, dt \geq \text{ (or } \leq\text{) } 0 \tag{72}$$

where $\delta x$ is an arbitrary variation of $x$ which is zero at $a$ and $b$. For $\delta A_k \geq 0$ (or $\leq 0$), we get

$$\frac{\partial}{\partial t}\left(\frac{\partial L_k}{\partial \dot{x}}\right) - \frac{\partial L_k}{\partial x} \geq \text{ (or } \leq\text{) } 0 \tag{73}$$

which can be proved by contradiction as follows. Suppose $\int_a^b f(t) \delta x \, dt \geq 0$ and $f(t) \leq c \leq 0$ during a small period of time $\Delta t$ somewhere between $a$ and $b$. Since $\delta x$ is arbitrary, let it be zero outside $\Delta t$ and a positive constant within $\Delta t$. We clearly have $\int_a^b f(t) \delta x \, dt \leq c \delta x \leq 0$, which contradicts our starting assumption. This proves Eq.(73).

The Legendre transformation $H_k = P_k \dot{x} - L_k$ along a path $k$ implies the momentum given by $P_k = \frac{\partial L_k}{\partial \dot{x}}$ which can be put into Eq.(73) to have

$$\dot{P}_k \geq (or \leq) \frac{\partial L_k}{\partial x} \tag{74}$$

for $\delta A_k \geq (or \leq) 0$.

However, from the path average of Eq.(72) and the SAP $\overline{\delta A} = 0$, we straightforwardly write

$$\overline{\delta A} = \sum_k p_k \int_a^b \left[ \frac{\partial}{\partial t}\left(\frac{\partial L_k}{\partial \dot{x}}\right) - \frac{\partial L_k}{\partial x} \right] \delta x \, dt = \int_a^b \left[ \overline{\frac{\partial}{\partial t}\left(\frac{\partial L_k}{\partial \dot{x}}\right)} - \overline{\frac{\partial L_k}{\partial x}} \right] \delta x \, dt = 0 \tag{75}$$

which implies

$$\overline{\frac{\partial}{\partial t}\left(\frac{\partial L_k}{\partial \dot{x}}\right)} - \overline{\frac{\partial L_k}{\partial x}} = 0. \tag{76}$$

This is the Euler-Lagrange equation of the random dynamics. We have equivalently

$$\overline{\dot{P}} = \overline{\frac{\partial L}{\partial x}}. \tag{77}$$

where $\overline{\dot{P}} = \sum_k p_k \dot{P}_k$ and $L = \sum_k p_k L_k$. This is just the average Newtonian second law mentioned in Eq.(10).

b) *Hamiltonian equations*

From Legendre transformation, we can have $\frac{\partial L_k}{\partial x} = -\frac{\partial H_k}{\partial x}$ and the following Hamiltonian equations

$$\dot{x}_k = \frac{\partial H_k}{\partial P_k} \text{ and } \dot{P}_k \geq (or \leq) -\frac{\partial H_k}{\partial x} \tag{78}$$

For a path $k$ along which $\delta A_k \geq (or \leq) 0$.

Naturally, Eq.(76) means

$$\overline{\dot{P}} = \overline{\frac{\partial H}{\partial x}}. \tag{79}$$

with the average Hamiltonian $H = \sum_k p_k H_k$.

## 17) Liouville theorem again

Now we are entitled to review Liouville theorem already discussed for regular dynamics in section 3. What will happen when the dynamics is random?

From Eq.(7), we know that $\left.\dfrac{d\rho}{dt}\right|_k = -\left(\dfrac{\partial \dot{x}_k}{\partial x} + \dfrac{\partial \dot{P}_k}{\partial P_k}\right)\rho$ along a path $k$. Let us write the second equation of the Eqs.(78) à la Langevin $\dot{P}_k = -\dfrac{\partial H_k}{\partial x} + R_k$ where $R_k \geq (or \leq) 0$ for $\delta A_k \geq (or \leq) 0$ is the random force causing the deviation from Newtonian laws. The average Newtonian law $\overline{\dot{P}} = \overline{\dfrac{\partial H}{\partial x}}$ yields $\sum_k p_k R_k = 0$ for any time moment of the process. We have

$$\left.\dfrac{d\rho}{dt}\right|_k = -\dfrac{\partial R_k}{\partial P_k}\rho . \tag{80}$$

and

$$\dfrac{d\rho}{dt} = \sum_k p_k \left.\dfrac{d\rho}{dt}\right|_k = -\overline{\dfrac{\partial R_k}{\partial P_k}}\rho . \tag{81}$$

Where $\overline{\dfrac{\partial R_k}{\partial P_k}} = \sum_k p_k \dfrac{\partial R_k}{\partial P_k}$ is an average over all the possible paths. The solution of this equation is

$$\rho(t) = \rho(t_0)\exp[\zeta(t,t_0)] \tag{82}$$

where the function $\zeta(t,t_0) = -\int_{t_0}^{t} \overline{\dfrac{\partial R_k}{\partial P_k}} dt$. The relationship $\sum_k p_k R_k = 0$ implies $\overline{\dfrac{\partial R_k}{\partial P_k}} = -\sum_k R_k \dfrac{\partial p_k}{\partial P_k}$ and, by using the exponential distribution of action, it can be proved $\dfrac{\partial p_k}{\partial P_k} = -\gamma p_k dx_k$, where $dx_k$ is a true (not virtual) displacement of the motion at time $t$ along the path $k$. Hence $\overline{\dfrac{\partial R_k}{\partial P_k}} = \gamma \sum_k p_k R_k dx_k = \gamma \overline{\delta W_R(t)}$ where $\overline{\delta W_R(t)} = \sum_k p_k dW_{Rk}$ is the average of the random work $dW_{Rk} = R_k dx_k$ performed by the random forces over the random displacement $dx_k$ along the path $k$ at time $t$. Finally, $\zeta(t,t_0) = -\gamma \int_{t_0}^{t} \overline{\delta W_R(t)} dt = -\gamma W_R(t,t_0)$ here $W_R(t,t_0) = \int_{t_0}^{t} \overline{\delta W_R(t)} dt$ is the cumulate average work of random force performed from $t_0$ and $t$,

and

$$\rho(t) = \rho(t_0)\exp[-\gamma W_R(t,t_0)] \tag{83}$$

meaning that the state density decreases (increases) and the phase volume increases (decreases) whenever the cumulate random work $W_R(t,t_0) > 0$ ($<0$). $\rho(t)$ is constant only when there is no work of random forces.

If the average random work $\overline{\delta W_R(t)}$ does not depend on time, we have $W_R(t,t_0) = \overline{\delta W_R}(t-t_0)$ and

$$\rho(t) = \rho(0)\exp[-\lambda t] \tag{84}$$

where $\lambda = \gamma \overline{\delta W_R}$ is a Lyapunov-like exponent characterizing the variation of the distances between the state points.

The phase volume $\Omega$ can be calculated by

$$\Omega_t = \int_\Gamma d\Gamma = \int_\Gamma \frac{dn}{\rho(t)} = \Omega_0 \exp[\lambda t] \tag{85}$$

where $dn$ is the number of state points in the phase volume $d\Gamma$ and $\Omega_0$ is the initial phase volume $\Omega_0 = \int_\Gamma \frac{1}{\rho(0)} dn$. The variation rate of phase volume is given by

$$\frac{d\Omega_t}{dt} = \Omega_0 \lambda \exp[\lambda t] \tag{86}$$

Where $\Omega_0$ is the initial phase volume and the volume $\Omega = \int_\Gamma \frac{1}{\rho} dn$ should be of course calculated in the occupied volume in which $dn \neq 0$.

The probability distribution of states $p(x,P,t)$ is proportional to the state density and reads

$$p(x,P,t) = p(x_0,P_0)\exp[-\lambda t] \tag{87}$$

If $p(x,P,t)$ at the time $t$ and $p(x_0,P_0)$ at $t=0$ are two equilibrium states described by Boltzmann distribution and Gibbs-Shannon formula $S = -\sum_{x,P} p(x,P)\ln p(x,P)$ for entropy, it is straightforward to see that the entropy production rate during the process is $\dot{S} = \dfrac{S(t)-S(0)}{t} = \lambda$. The same result can be reached if we use the Boltzmann definition of entropy $S = \ln\Omega$.

An application of this scheme to gas expansion may be instructive for understanding the microscopic physics why entropy increases in the process. If a gas evolves from an equilibrium state at time $t=0$ into another equilibrium state at time $t$, and if the entropy of the two equilibrium state can be given by the Gibbs-Shannon formula, then from Eqs.(42) and (77), we get $\Delta S = N\ln\dfrac{V_2}{V_1} = -\zeta(t,t_0) = \gamma W_R(t,t_0)$ (Boltzmann constant $k_B=1$). This means that the entropy increase is a consequence of the work performed by the random forces. This also implies that the cumulate work of random forces must vanish in equilibrium system. In other words, the work of random force is not zero only when there are nonequilibrium processes. This sounds plausible.

18) Poincaré recurrence theorem again

From the above results, if the average cumulate work of random forces $W_R(t,t_0)$ is not zero, the phase volume varies in time such that

$$\Omega_t = \Omega_0 \exp[\gamma W_R(t,t_0)] \tag{88}$$

When the cumulate work is negative or positive, $\Omega$ is decreasing or increasing, respectively. $\Omega$ is conservative only when the cumulate work is vanishing.

As a consequence, the Poincaré recurrence theorem is in general lost, since $\Omega$ is no more conserved by random dynamics. In the case of increasing entropy, $\Omega$ is increasing according to Eq.(82). For example, in the case of isolated gas expansion, energy is conserved and entropy is increasing, implying that $\gamma W_R(t,t_0) \geq 0$. If the initial and final states are in equilibrium, $\gamma W_R(t,t_0) = N\ln\dfrac{V_2}{V_1}$ and

$$\Omega_t = \Omega_0 \left(\dfrac{V_2}{V_1}\right)^N. \tag{89}$$

Clearly, there is no Poincaré recurrence in the present formalism of random dynamics.

19) Uncertainty relationships of the random dynamics

At this stage, it is worth reviewing the result of a previous work[22] about a commutation relationship and the concomitant uncertainty relationship of the random dynamics described by the path probability $p_k \propto e^{-\gamma A_k(a,b)}$. We can define a function of path $G_k(x)$ of which the average over all the path by $\overline{G} = \sum_k p_k G_k(x)$ or by path integral[39]

$$\overline{G} = \int_{x_a}^{x_b} dx_1 \int_{x_a}^{x_b} dx_2 ... \int_{x_a}^{x_b} dx_n \, p_k(x_a...x_b) G_k(x_a...x_b). \tag{90}$$

A subsequent relationship is[22][39]

$$\overline{\frac{\partial G_k(x)}{\partial x}} = -\gamma \overline{G_k(x) \frac{\partial A_k(a,b)}{\partial x}}. \tag{91}$$

If we choose $G_k(x) = x$, it follows

$$\overline{P_{i+1} x_i} - \overline{x_i P_i} = \frac{1}{\gamma} \quad \text{or} \quad \overline{[P_{i+1}, x_i]} = \frac{1}{\gamma} \tag{92}$$

where $i$ is the index of a segment of a path $k$ corresponding to the time period $\delta t = t_{i+1} - t_i$, $x_i$ is the position at time $t_i$, $P_{i+1} = \frac{x_{i+1} - x_i}{\delta t}$ and $P_i = \frac{x_i - x_{i-1}}{\delta t}$ are the momentum at time $t_{i+1}$ and $t_i$, respectively. From the experimental point of view (see for example section 8), the observations at the three time moments $t_{i+1}$, $t_i$ and $t_{i+1}$ are inevitable for determining the system's behavior of the moment $t_i$, the quantities such as position and velocity at these moments are correlated or associated one with another. There is no sense to distinguish them by the indices of subsequent moments of time. With this in mind, it makes sense to write, for a given moment $t_i$

$$\overline{Px} - \overline{xP} = \frac{1}{\gamma} \quad \text{or} \quad \overline{[P,x]} = \frac{1}{\gamma}. \tag{93}$$

Now let us define $\Delta x_k = x_k - \overline{x}$, $\Delta P_k = P_k - \overline{P}$ and $C_k = \alpha \Delta P_k - \Delta x_k$. It is straightforward to calculate $\overline{C_k^2} = \alpha^2 \overline{\Delta P_k^2} + \alpha \frac{1}{\gamma} + \overline{\Delta x_k^2} \geq 0$. This means $\frac{1}{\gamma^2} - 4 \overline{\Delta P_k^2} \cdot \overline{\Delta x_k^2} \leq 0$ hence

$$\overline{\Delta x^2} \cdot \overline{\Delta P^2} \geq \frac{1}{4\gamma^2}. \tag{94}$$

A similar relation (without the factor ¼ at the right hand side) has been found in our previous work [22] with a method of path integral. With the above approach, if we define $C_k = \alpha \Delta A_k - 1$, we will find $\overline{\Delta A^2} \geq \frac{1}{4\gamma^2}$, $\overline{\Delta L^2} \cdot \overline{\Delta t^2} \geq \frac{1}{4\gamma^2}$ and $\overline{\Delta H^2} \cdot \overline{\Delta t^2} \geq \frac{1}{4\gamma^2}$ as in [22].

The physical signification of these relations is still a matter of investigation. Our understanding is that, due to the randomness of the motion, the random variable action has a minimal experimental error $\frac{1}{4\gamma^2}$, hence all the pairs of conjugate variables ($P$ and $x$, $L$ and $t$, as well as $H$ and $t$) in the definition of action ( $A = \int_a^b L dt = \int_a^b P dx - \int_a^b H dt$ ) are (jointly) affected. However, since these uncertainty relations are statistical results averaged over all the possible paths, it does not have restriction on the single measurement, meaning that a priori the momentum and position can be measured in a single experiment with arbitrary independent precision. To our (current) opinion, at least at the present stage, these relations are not laden with the same physical and philosophical contents as the quantum uncertainty relations (complementarity, entanglement, perturbation of observers etc). But this conceptual aspect of the is still a matter of investigation. The study of Brownian entanglement[45] is an example, although the authors considered the uncertainty relation[46] relating position $x$ and drift (osmotic) velocity $u$ instead of the momentum $P$ we used here.

In any case, if the dynamical randomness is vanishing, $\gamma \to \infty$ ($D \to 0$), these relations are no more significant since, for example, $\overline{\Delta x^2} \cdot \overline{\Delta P^2} \geq 0$.

Another remark is that the relation $\overline{\Delta x^2} \cdot \overline{\Delta P^2} \geq \frac{1}{4\gamma^2}$ will allow us to avoid the chimerical relation in the density of state usually calculated by, say, $dn = \frac{dxdP}{\hbar}$ for a two dimensional phase space. It will be possible to replace it by $dn = \frac{dxdP}{1/\gamma}$ to avoid the quantum constant in classical mechanics. This is more comfortable especially when the treated system and its components are non quantum bodies.

The value of the uncertainty can be estimated by the formula $1/\gamma = 2mD$ and $D = k_B T / 6\pi\eta r$ for Brownian motion, where $k_B$ is the Boltzmann constant, $T$ is the absolute temperature (we choose $T$=300K), $\eta$ the viscosity of the solvent (we choose water $\eta$=0.001

kg/m.s), $r$ the radius of the Brownian particle (we choose a bacteria with $r \approx 750$ nm), and $m \approx 10^{-16}$ kg the mass of a bacteria. It follows that $D \approx 3 \times 10^{-13}$ m$^2$/s and $1/\gamma = 2mD \approx 6 \times 10^{-29}$ J.s.

For a fullerene $C_{60}$ in water with $r \approx 1$ nm and $m \approx 1.2 \times 10^{-24}$ kg, we get $D \approx 2.3 \times 10^{-10}$ m$^2$/s and $1/\gamma \approx 5.5 \times 10^{-34}$ J.s. This value is comparable to the Planck constant.

## 20) Concluding remarks

To sum up in few words, by considering an experiment to verify Newton's second law and by introducing random trajectory to characterize the dynamical randomness, the virtual work principle and the least action principle are extended to random dynamics with $\overline{\delta W} = 0$ and $\overline{\delta A} = 0$. A probabilistic mechanics theory is formulated from these two principles.

The main results are the following:

1) For equilibrium thermodynamic open system, it is proved that $\overline{\delta W} = -\delta \overline{E} + \mu \delta \overline{N} + \frac{\delta S}{\beta}$, Hence the maximum entropy algorithm $\delta(S - \beta \overline{E} + \beta \mu \overline{N}) = 0$ follows from the stochastic virtual work principle. The outcomes along this line are: a) Maximum entropy (maxent) is kind of dynamical equilibrium prescribed by the stochastic virtual work principle. b) This maxent does not need any assumption about the entropy property and functional form. c) The equiprobability for microcanonical ensemble is a natural consequence of this approach whatever the entropy form. d) The constraints introduced as partial knowledge into maxent on the basis of informational arguments appear here naturally as a consequence of vanishing virtual work.

2) For nonequilibrium systems, we obtained a relation $\delta S = \delta S^{eq} + \delta S^{neq}$. It is proved that in certain specific cases (heat conduction, free expansion of ideal gas etc) the variation $\Delta S^{neq}$ between two equilibrium states is equal to the change of entropy.

3) It is shown that, in steady process with local equilibrium, $\delta S^{neq}$ can be given by the well known formula of entropy production rate $\dot{S}^{neq} = \sum_i \vec{F}_i \cdot \vec{J}_i$ where $\vec{F}_i$ is a thermodynamic force and $\vec{J}_i$ the corresponding transport flux.

4) $S^{neq}$ has a extremum through the algorithm $\delta[S^{neq} - \eta(\overline{E} - \overline{W_\lambda})] = 0$ for any nonequilibrium process driven by thermodynamic forces and an external force controlled through a parameter $\lambda$.

5) In case where the path information is given by Shannon formula, the path probability is given by $p_k \propto e^{-\gamma A_k(a,b)}$ between two points $a$ and $b$ in configuration space. This path probability satisfies a Fokker-Plank equation for normal diffusion[23].

6) The Euler-Lagrange equation of this random dynamics along any path is now

$$\frac{\partial}{\partial t}\left(\frac{\partial L_k}{\partial \dot{x}}\right) - \frac{\partial L_k}{\partial x} \geq \text{(or } \leq\text{) } 0 \text{ instead of } \frac{\partial}{\partial t}\left(\frac{\partial L_k}{\partial \dot{x}}\right) - \frac{\partial L_k}{\partial x} = 0.$$

7) The random dynamics described by the path probability $p_k \propto e^{-\gamma A_k(a,b)}$ has a commutation relation $\overline{Px} - \overline{xP} = \frac{1}{\gamma}$ and the following uncertainty relations

$$\overline{\Delta A^2} \geq \frac{1}{2\gamma^2}, \quad \overline{\Delta x^2} \cdot \overline{\Delta P^2} \geq \frac{1}{\gamma^2}, \text{ and } \overline{\Delta H^2} \cdot \overline{\Delta t^2} \geq \frac{1}{2\gamma^2}.$$

8) The conventional Liouville theorem is violated. It has been demonstrated that the phase space ($\Gamma$) distribution function $\rho(\Gamma,t)$ evolves in time according to $\rho(\Gamma,t) = \rho(\Gamma_0,0)\exp[-\lambda t]$ with $\lambda = \gamma \overline{\delta W_R}$ and $\overline{\delta W_R}$ is the work of random forces averaged over all the paths in the period $0 \leq \tau \leq t$. In term of phase volume $\Omega$, the theorem reads $\Omega_t = \Omega_0 \exp[\lambda t]$. $\lambda$ is the entropy production rate between two equilibrium states at τ=0 and τ=t, respectively. The Poincaré recurrence theorem is no more valid.

Some other remarks about this formalism :

1) An advantage of this approach is the quantitative treatment of entropy variation of nonequilibrium system without extending entropy notion to nonequilibrium states. It is our hope that this may be helpful for the establishment of a nonequilibrium thermodynamics and statistical mechanics without the hypothesis of local equilibrium.

2) Justification of the variational methods in both equilibrium and nonequilibrium statistical mechanics on the basis of the basic principles of mechanics. One can definitely say for example that the maximum entropy method is not only an inference approach relevant to the subjective nature of probability, but a physical law on its own.

3) The utmost aim of this approach is to give an answer (perhaps not the only one) to the longstanding enigma of the emergence of time arrow from the underlying time reversible dynamics. The time arrow emerges from an underlying dynamics which is intrinsically irreversible due to the random forces. But the final aim is not yet reached. For the time being, we have proved for some cases (free expansion of gas, heat conduction) that the nonequilibrium uncertainty $S^{neq}$ increases and $\Delta S^{neq}$ equals the entropy change between two equilibrium states, but there is not yet general proof saying that $\Delta S^{neq}$ is always positive for any isolated system. The search for this proof will be one of the main tasks in the future. Another equivalent work is to prove the positive random work $W_R(t,t_0) \geq 0$ for any isolated system.

4) If the above work is successful, the second law will read $\Delta S^{neq} \geq 0$ if the system is isolated. In view of the equation $\Delta S = \Delta S^{neq} + \Delta S^{eq}$ for the total uncertainty $S$ and the fact that $\Delta S^{eq} = \dfrac{\Delta Q}{T}$ for reversible process, the second law can be written as $\Delta S \geq \sum_i \dfrac{\Delta Q_i}{T_i}$ with a slightly different signification from the original Clausius inequality, since the "$S$" is not the entropy defined for equilibrium state. This inequality should be valid for any process involving heat exchanges $\Delta Q_i$ at temperatures $T_i$ during the process.

5) Our final remark is that this approach raises a question about the use of probability description in regular and deterministic dynamics. An example of this use is the discussion of entropy variation through Liouville theorem with the probability distribution of states proportional to the state density in phase space of mechanical system. Yet in regular dynamics, each trajectory in phase space can be a priori traced in time with certainty, each state has its unique time to be visited by the system, and the frequency of visit by the system of any phase volume can be predicted exactly. Hence *in principle* there is no place for probability and entropy or other uncertainty. If it is necessary *in practice* to use probability description due to the very large number of degrees of freedom which overpass our capacity of mathematical and numerical description, this probability is not an intrinsic property of the dynamics. Consequently, the use of this pragmatic probability in the discussion of entropy with the Liouville theorem is questionable. Intrinsic probability exists only when the dynamics is probabilistic meaning that not only an observer does not know where a system is going at next time step, the system itself also "ignores" it. This unpredictability is out of the realm of regular dynamics.